\documentclass{jfm}
\usepackage{amsmath}
\usepackage{graphicx}
\usepackage{lipsum} 
\usepackage{caption}
\usepackage{physics}
\usepackage{hyperref}
\usepackage{newtxmath}
\usepackage{amsmath}
\usepackage{soul}
\usepackage{amssymb}
\usepackage{textcomp, gensymb}
\usepackage{comment}
\usepackage{booktabs}
\usepackage{xcolor}
\usepackage{natbib}
\usepackage{float}
\shorttitle{ventilated cavities}
\shortauthor{Gawandalkar U.U. and Lucido L.A. et al.}

\title{Examination of ventilated cavities in the wake of a two-dimensional bluff body}

\author{Udhav U. Gawandalkar\aff{1}
  \corresp{\email{u.u.gawandalkar@tudelft.nl}},
  Nicholas A. Lucido\aff{2}, Prachet Jain\aff{2}, Christian Poelma\aff{1},   Steven L. Ceccio\aff{2,3}, Harish Ganesh\aff{2}
 }

\affiliation{\aff{1}Process and Energy, Mechanical Engineering, Delft University of Technology, Leeghwaterstraat 39, 2628 CB Delft, The Netherlands
\aff{2}Department of Naval Architecture and Marine Engineering, University of Michigan, Ann Arbor, MI, USA
\aff{3} Department of Mechanical Engineering, University of Michigan, Ann Arbor, MI, USA
}

\begin{document}

\maketitle

\begin{abstract}
%stable cavity part
\textcolor{black}{Ventilated cavities in the wake of a two-dimensional bluff body are studied experimentally via time-resolved X-ray densitometry. With a systematic variation of flow velocity and gas injection rate, expressed as Froude number ($Fr$) and ventilation coefficient ($C_{qs}$), four stable cavities with different closures are identified. A regime map governed by $Fr$ and $C_{qs}$ is constructed to estimate flow conditions associated with each cavity closure type. Each closure exhibits a different gas ejection mechanism, which in turn dictates the cavity geometry and the pressure inside the cavity. Three-dimensional cavity closure is seen to exist for the supercavities at low $Fr$. However, closure is nominally two-dimensional for supercavities at higher $Fr$. At low $C_{qs}$, cavity closure is seen to be wake-dominated, while supercavities are seen to have wave--type closure at higher $C_{qs}$, irrespective of $Fr$.
%unstable cavity part
With the measured gas fraction, a simple gas balance analysis is performed to quantify the gas ejection rate at the transitional cavity closure during its formation. 
For a range of $Fr$, the transitional cavity closure is seen to be characterised by liquid re-entrant flow, whose intensity depends on the flow inertia, dictating the gas ejection rates.  
Two different ventilation strategies were employed to systematically investigate the formation and maintenance gas fluxes.
The interaction of wake and gas injection is suspected to dominate the cavity formation process and not the maintenance, resulting in ventilation hysteresis. Consequently, the ventilation gas flux required to maintain the supercavity is significantly less than the gas flux required to form the supercavity. }
\end{abstract}


% ============================ Introduction ============================ %
\section{Introduction}\label{sec:introduction}
Ventilated partial-- and super--cavitation (VPC and VSC, respectively) are characterised by gas cavities formed by injecting non-condensable gas behind a `cavitator' \citep{Logvinovich1969}. This technique has gained significant attention due to its potential application for drag reduction, for instance on ship hulls by forming an air cavity and reducing the near-wall density \citep{Ceccio2010}. Ventilated supercavities have also found applications in hydraulic engineering \citep{Chanson2010} and process industries \citep{Rigby1997} to mitigate deleterious effects of natural cavitation, such as wear, erosion, and failure, all resulting from violent cloud implosion \citep{Brennen1995}. The stability of ventilated cavities is crucial, as unstable ventilated cavities can get detached abruptly, leading to a sudden increase in drag forces. Insufficient ventilation may lead to cavity collapse, while excessive ventilation could result in cavity oscillation, both undesired and often detrimental \citep{Ceccio2010}. Hence, it is necessary to understand the exact flow conditions that govern the stability of ventilated cavities. Ventilated cavities are governed by incoming flow pressure ($P_{0}$) and velocity ($U_{0}$), input gas injection rate ($\dot Q_{in}$), cavity pressure ($P_{c}$), and cavitator geometry (area $A=WH$, length scale $H$). See figure \ref{fig:setup} for the definitions of parameters. These parameters can be expressed as non-dimensional cavitation number ($\sigma_{c}$), Froude number ($Fr$), and ventilation coefficient ($C_{qs}$) as defined below. 
\begin{equation*} 
    \sigma_{c} = \frac{P_{0}-P_{c}}{\frac{1}{2} \rho U_{0}^{2}} \hspace{7mm}  Fr = \frac{U_{0}}{\sqrt{gH}} \hspace{7mm} C_{qs} = \frac{\dot Q_{in}}{U_{0}A}.
\end{equation*}
Ventilated cavities are formed when a part of the injected gas ($\dot Q_{in}$) gets entrained in the separated flow (behind the cavitator), while the remainder of the gas is ejected ($\dot Q_{out}$) from the cavity closure region. The entrained gas, i.e. the gas that is dragged \emph{into} the cavity results in the growth of the cavity. Here, closure refers to the way a cavity closes itself and dictates the amount of gas ejected out of the cavity. The cavity closure also influences the cavity geometry (length, thickness and gas distribution), and, most importantly, the stability of the cavity. Hence, a thorough understanding of the cavity closure is imperative. 

Various closure types have been observed in the past, especially three-dimensional axisymmetric cavitators have been investigated widely. It was shown that at a high $Fr$, the cavity closure is characterised by a weak re-entrant jet \citep{EpshteynLA1961, Logvinovich1969, Karn2016}. However, at low $Fr$, buoyancy effects result in lift generation and the formation of two vortex tubes at the closure \citep{semen}. \cite{Karn2016} explained these observations based on the pressure difference across the cavity closure ($\Delta \tilde{P}$): a higher $\Delta \tilde{P}$ gave rise to a re-entrant jet similar to natural partial cavities \citep{Knapp1958, Callenaere2001}, while a lower $\Delta \tilde{P}$ resulted in vortex tube closure. At significantly high ventilation inputs, oscillating cavities called pulsating cavities (PC) were identified \citep{silbersong61, skid}. For the three-dimensional fence-type cavitator, 
similar observations were made: cavities with re-entrant flow were seen at higher $Fr$, while at lower $Fr$ the cavity was seen to split into two separate branches with re-entrant flow on each branch \citep{Barbaca2017}. All the cavities had a re-entrant flow closure, possibly due to the high Froude number ($Fr$) employed in that study. Ventilated cavities behind a two-dimensional cavitator have received less attention in the literature despite their wide application for partial cavity drag reduction on ships \citep{Makiharju2013a}.
In a wall-bounded two-dimensional cavitator, \cite{Qin2019} observed a twin-branch cavity (TBC), analogous to vortex tube closure. \cite{Qin2019} also reported supercavities with dispersed bubbles at the closure, and cavities with re-entrant jet closure were \emph{not} observed, likely due to the low Froude number ($Fr<6$) considered in their study. In summary, distinct closure types are observed for different cavitator geometries. \textcolor{black}{To this end, ventilated cavities in the wake of two-dimensional bluff bodies remain rather unexplored.}

The cavity closure influences the gas entrainment/ejection rate into/out of the cavity. Understanding the gas entrainment and ejection mechanisms is important to establish and maintain ventilated cavities efficiently. \cite{Spurk2002} postulated that the injected gas is carried to the closure by a growing internal boundary layer at the gas--liquid interface, where it is ejected out in the form of toroidal vortices. For wall-bounded cavitators, \cite{Qin2019} proposed that the re-circulation region interface is responsible for entraining the gas bubbles into the separated shear layer, which are then carried away downstream. This was verified experimentally by \cite{Wu2019a}, in a study on the gas flow \emph{inside} the ventilated cavity using particle image velocimetry (PIV). Although the gas entrainment mechanisms were found to be identical for different cavity closures, the gas leakage mechanisms were seen to be different. Various gas ejection mechanisms are identified in ventilated cavities; (i)  gas ejection due to a re-entrant jet \citep{Spurk2002, Kinzel2010, Barbaca2017}, (ii) vortex tube gas leakage \citep{semen, Cox1955}, (iii) pulsation of cavities \citep{Michel1984, skid, Karn2016}, and (iv) surface waves pinching the cavity \citep{Zverkhovskyi}. Furthermore, \cite{Qin2019} observed the role of capillary wave pinch-off in gas ejection in twin-branched cavities (TBC).

\textcolor{black}{Despite qualitative observations,} studies dedicated to the quantification of gas ejection rates out of cavities ($\dot{Q}_{out}$) for different cavity closures are scarce. Yet, quantifying gas ejection rates is essential in formulating empirical models, validating numerical models, and furthering our understanding of the underlying flow physics at cavity closures. This also enables better prediction of gas ventilation demands under different flow conditions ($Fr$ and $C_{qs}$). Recently, \cite{Shao2022} used digital inline holography (DIH) to quantify the instantaneous $\dot{Q}_{out}$ for stable cavity closure types. While stable cavity types have received adequate attention in the literature, transitional cavity closure types during the formation of a supercavity remain unexplored. The ventilation demands to establish and maintain VCs can be more accurately estimated by studying the gas ejection of these transitional VCs. Furthermore, hysteresis in VC formation plays a significant role in determining the accurate ventilation demands, i.e. for a given ventilation ($C_{qs}$), a ventilated cavity can assume a different length and closure depending upon how the ventilation condition was reached \citep{Karn2016, Kawakami2011, Makiharju2013a}. Ventilation hysteresis is widely reported in ventilated cavities, but the exact physical mechanism responsible for it remains unclear. The characterisation and implication of ventilation hysteresis are essential to devise control strategies for efficient drag-reduction and aeration systems.

The lack of quantitative insights in ventilated cavities can be attributed to the challenges brought about by turbulence, frothiness, and optical opaqueness of the flow. Thus, conventional optical-based measurement techniques are untenable, especially at the liquid-gas-liquid interface and cavity closure region. \cite{Wosnik2013} attempted to estimate the void fraction and velocity fields in the frothy mixture of the VCs with laser-illuminated bubble images, however, the uncertainty in the measurements was high. Holography is limited to the far-field, where individual ejected gas bubbles can be imaged \citep{Shao2022}. High-fidelity numerical simulations like DNS are limited to low Reynolds numbers \citep{Liu2023} due to the large density ratios and turbulent motions, with a wide range of scales in the flow \citep{Madabhushi2023}. These shortcomings can be overcome by whole-field radiation-based measurement techniques such as time-resolved X-ray densitometry \citep{Aliseda2021, Makiharju2013}, wherein gas-liquid interfaces and the cavity closure region can be resolved reliably. Such time-resolved void fraction measurements can provide quantitative information in gas entrainment and leakage dynamics apart from time-averaged gas distribution in ventilated cavities. The void fraction fields are also indispensable for quantifying the compressibility effects in ventilated cavity flows, deemed crucial in {\em{natural}} cavitation flows \citep{Ganesh2016, Gawandalkar2024}. Further, void fraction profiles in ventilated cavities can be useful for validating numerical models aimed at accurately simulating complex ventilated cavity flows. 

In this paper, we study the entrainment of gas in the wake of a two-dimensional wedge by systematically varying the flow inertia ($Fr$) and gas injection rate ($C_{qs}$). We identify four different types of stable ventilated cavities and determine the associated flow conditions on a regime map. \textcolor{black}{The effect of $Fr$ and $C_{qs}$ on the cavity closure type, cavity geometry, and gas ejection mechanism is studied in detail using two-dimensional time-resolved X-ray densitometry and high-speed imaging. The dynamics of supercavity formation process are studied for a range of $Fr$. The transitional cavity closures and the resulting gas ejection rates out of the cavity are quantified during the formation using a simple gas balance based on a control volume approach.} The ventilation hysteresis in the formation of supercavities is investigated systematically. We observe a substantial difference between the gas flux required to form and maintain a supercavity. \textcolor{black}{Finally, we link the above observation to the cavity closure and the resulting gas ejection rates influenced by wake-gas interaction.} The rest of the paper is organised into five additional sections. The experimental methodology is described in $\S$2. The stable and transitional ventilated cavities are treated separately; the characteristics of stable ventilated cavities are described in detail in $\S$3, while transitional cavities during the formation of supercavities are examined in $\S$4. Ventilation hysteresis in supercavity formation is discussed in $\S$5, followed by conclusions which are summarised in $\S$6. 

% ============================ Experimental methodology ============================ %
\section{Experimental methodology}
\label{exper}
\subsection{Flow setup}

\begin{figure}
    \centering
    \includegraphics[width=0.98\textwidth]{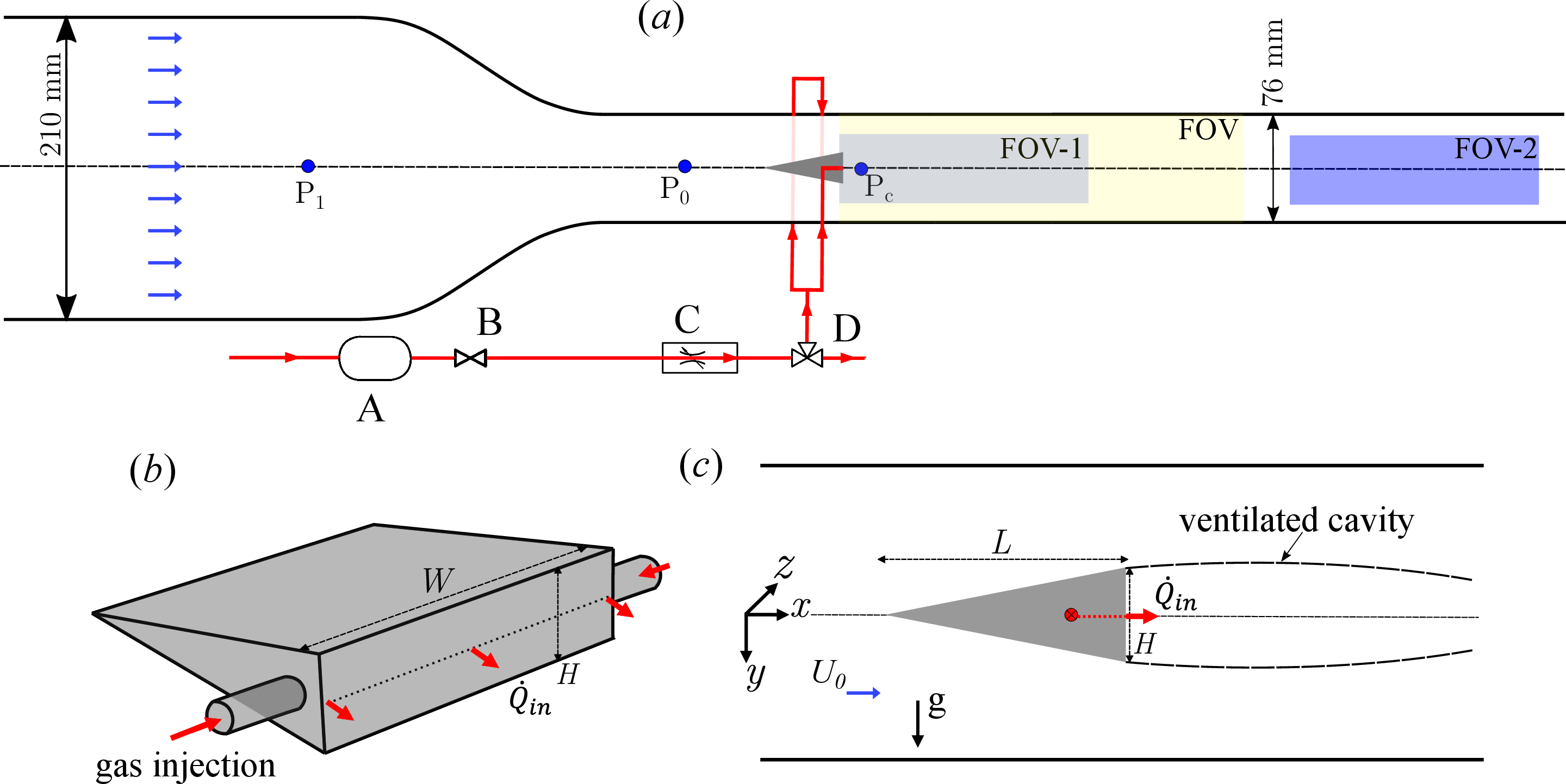}
    \caption{(\textit{a}) A schematic of the experimental flow facility with gas ventilation line shown in red, A: pressure vessel, B: valve, C: voltage regulated mass flow controller, D: three-way valve. The yellow region indicates the field of view (FOV) for high-speed optical imaging, while the grey (FOV-1) and blue regions (FOV-2) indicate the field of view of X-ray imaging. The pressure is measured at $P_{0}, P_{1}$, and $P_{c}$. (\textit{b}) A schematic of the wedge with ventilation holes. (\textit{c}) A schematic of the  ventilated cavity. Note that the red arrow shows the direction of the gas flow, while the blue arrow shows the direction of the bulk flow.}
    \label{fig:setup} 
\end{figure}

The experiments were performed at the University of Michigan in the 9-inch ($\approx$ 210 mm) recirculating water tunnel with a reduced square test-section of cross-section 76$\times$76 mm$^{2}$, as discussed by  \cite{Ganesh2016}. The inflow velocity in the test-section ($U_{0}$) was measured based on the pressure drop across the contraction ($\Delta P = P_{1} - P_{0}$, see figure \ref{fig:setup}(\textit{a})) using a differential pressure transducer (Omega Engineering PX20-030A5V). %between the contraction and the inlet of the test-section 
Inflow static pressure ($P_0$) was measured using an Omega Engineering PX409030DWU10V, 0 to 208 kPa transducer.  The experiments were performed at ambient system pressure, i.e. without any vacuum. Dissolved gas content was controlled using a deaeration system. The flow velocity ($U_{0}$) was varied from 0.84 ms$^{-1}$ to 6.2 ms$^{-1}$, corresponding to $Fr$ of 2 to 14 (see table \ref{Tab1} for definition).

The ventilated partial cavity was generated behind a two-dimensional wedge, by injecting non-condensable gas in its wake as shown in figure \ref{fig:setup}(\textit{c}). The wedge had a height ($H$) of 19 mm and an angle of 15$^{\circ}$, and was seal-secured tightly by the test-section windows, resulting in a blockage ($\xi$) of 25\%. The wedge had an internal bore-hole leading to multiple ventilation ports of 1 mm diameter each on the wedge base, as detailed in \cite{Wu2021}. The bore-hole was connected to the external compressed air supply via pneumatic fittings as shown in figure \ref{fig:setup}(\textit{b}). The gas ventilation line is schematically illustrated in figure \ref{fig:setup}(\textit{a}). The non-condensable gas was fed through a pressure vessel to maintain the required stagnation pressure ($\sim$ 400 kPa) to mitigate choking in the ventilation lines. The mass flow rate ($\dot{Q}_{in}$) was controlled using two flow controllers: Omega FMA series 0-15 and 0-50 standard litres per minute (SLPM). The gas injection rate was measured in SLPM due to a lack of pressure measurements during injection. The injected gas flow rate was expressed as non-dimensional ventilation coefficient $C_{qs}$ (see table \ref{Tab1} for definition) and was varied from 0.02 to 0.12 over more than 100 values for a given base pressure. The pressure inside the cavity ($P_{c}$) was measured from the side window of the test-section at $x \approx 1H$, along the wedge centreline, using an Omega Engineering PX409030DWU10V, 0 to 208 kPa transducer (see again  \ref{fig:setup}(\textit{a})). The measured cavity pressure, incoming pressure ($P_{0}$) and dynamic pressure of the incoming flow ($\frac{1}{2}\rho U_{0}^2$) were used to define the cavitation number expressed as $\sigma_{c}$ (see table \ref{Tab1} for definition). Note that cavitation number is used extensively in natural cavitating flows to indicate the closeness of cavity pressure to the vapour pressure \citep{Brennen1995}.

\begin{table} \setlength{\tabcolsep}{12pt} 
\begin{center}
\centering
\small
\begin{tabular}{*{4}{l}} \vspace{2mm}
Parameter & Definition & Units & Value \\ 
\hline
wedge dimensions & $L\times W \times H$ (see figure \ref{fig:setup})& mm&
 72$\times$76$\times$19 
\\ bulk inflow velocity & $U_{0}$  & ms$^{-1}$ & 0.84 - 6.2  \\
gas mass flow rate & $\dot{Q}$  & SLPM & 0.1 - 50 \\
Froude Number & $Fr = U_{0}/\sqrt{gH}$ & -- & 2 - 13.9 \\
ventilation coefficient & $C_{qs}= \dot{Q}/U_{0}HW$ & -- & 0.02-0.12  \\
cavitation number& $\sigma_{c} = (P_{0}-P_{c})/ \frac{1}{2}\rho U_{0}^{2}$ & -- & 0.8-3.5\\
Reynolds number & $Re_{H} = U_{0}H/\nu$ & -- & 1.6--11$\times$10$^{4}$\\
Strouhal number & $St_{H} = fH/U_0$ &  & \\
\end{tabular}\quad
\caption{Experimental parameters. $\rho$, $\nu$, and $g$ refer to the mass density, kinematic viscosity of water, and acceleration due to gravity, respectively.}
\label{Tab1}
\end{center}
\end{table}

\subsection{Flow visualisation}
Visual observation and qualitative analyses of ventilated partial cavities were performed via front-illuminated high-speed cinematography using a single Phantom Cinemag 2 v710 camera placed perpendicular to the field-of-view (FOV). The FOV was centered along the test-section axis and spans 13.6$H$$\times$8.4$H$ in the $x-y$ plane with the origin ({$x$, $y$, $z$} = 0) defined at the center of the wedge base. See the yellow region in figure \ref{fig:setup}(\textit{a}) for the FOV. \textcolor{black}{In a separate set of experiments, auxiliary high-speed visualisations were also performed to image a top-view of the cavity in $x-z$ plane with similar settings.} The camera was equipped with a 105 mm Nikkor lens set to $f^{\#}$ = 5.6 to allow sufficient contrast in images. The images were acquired at 500--2000 Hz for $\sim$ 11--44 seconds, depending on the nature of the experiment. Time-resolved, spanwise-averaged void fraction fields of ventilated cavities were measured using a high-speed two-dimensional X-ray densitometry system described in detail in \cite{Makiharju2013}. The current and the voltage of the X-ray source were set to 140 mA and 60 kV respectively, resulting in a measurement time of 1.6 seconds. The FOV-1, corresponding to X-ray densitometry, spanned 8.2$H$ $\times$ 4$H$ in the $x-y$ plane (see grey region figure \ref{fig:setup}\textit{a}). An obstruction in the line of sight of X-rays resulted in a small disc-shaped, nonphysical artefact in the void fraction fields, located at $x/H \approx$ 0.78, $y/H \approx$ -0.52: see for instance the black disc in figure \ref{fig:twin-cavity}(\textit{c}). The instantaneous void fractions were estimated with a spatial resolution of 0.16 mm ($0.0084H$) and a temporal resolution of 0.001 seconds. For long cavities with closure in the region beyond the X-ray measurement domain (FOV-1), only qualitative X-ray visualisation (i.e. no {\em{quantitative}} void fraction field measurements) could be performed from $x$ = 14$H$ to 22$H$. This alternate FOV is shown by the blue region marked `FOV-2' in figure \ref{fig:setup}(\textit{a}). The thicker and denser PVC walls of the test facility led to a substantial reduction in signal-to-noise ratio (SNR) precluding the quantification of void fractions.  The image acquisition (high-speed photography and X-ray imaging) was time-synchronised with gas ventilation input ($\dot{Q}$) and pressure transducers ($P_{1}$, $P_{0}$, $P_{c}$) using a digital pulse generator (DG535, Stanford Research Systems).

\subsection{Experimental procedure} \label{procedure}
\begin{figure}
    \centering
    \includegraphics[width=0.45\textwidth]{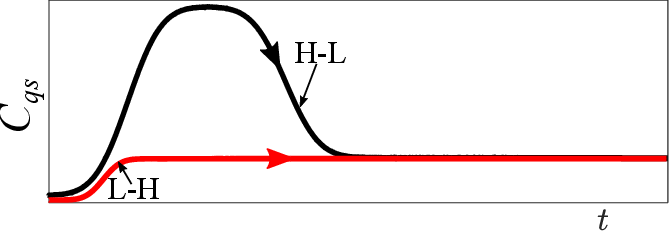}
    \caption{Gas injection profiles in time: red indicates a typical L-H profile, while black indicates a typical H-L profile. The grey region indicates the measurement time interval.}
    \label{fig:injection-profile} 
\end{figure}

The experiments were performed for a range of $U_{0}$ ($Fr$) and over 100 different ventilation inputs ($C_{qs}$). Since the aim of this study was also to examine the formation dynamics of a ventilated cavity, we performed two different types of experiments: In the first set of experiments, the gas was injected from no injection ($C_{qs} \sim$ 0) to the desired $C_{qs}$ with a prescribed error function, i.e. $(d{C_{qs}}/dt > 0$ or $\dot{C}_{qs}>0$). This ventilation strategy is referred to as `L-H', as the ventilation is increased from zero to a given $C_{qs}$ (see red profile in figure \ref{fig:injection-profile}) in about 5 seconds and kept constant for at least 10 seconds depending on the nature of the experiment. 
In the second set of experiments, a fully-developed supercavity was used as an initial condition, and the gas injection rate was reduced with an error function to achieve the desired $C_{qs}$, i.e. $ \dot{C}_{qs}<0$ (see black profile in figure \ref{fig:injection-profile}). This ventilation strategy is referred to as `H-L' as the ventilation rate is reduced. The high $C_{qs}$ is maintained for at least 5 seconds to ensure that the supercavity closure is fully developed before reducing it to the final low $C_{qs}$, which is kept constant for at least 15 seconds. Thus, \textcolor{black}{for both strategies,} upon establishing a cavity, measurements were performed after waiting for sufficient time to ensure that the cavity length did not change. \textcolor{black}{See the grey region in figure \ref{fig:injection-profile}.} Note that the effect of the rate of increases of $C_{qs}$ is beyond the scope of the current study. The ventilation was increased/decreased smoothly with an error function to mitigate the sharp overshoot in $\dot{Q}_{in}$ inherent to the first-order step response of the mass flow controller. This allowed precise control of the volume of gas injected in the flow. After each measurement, the flow loop was carefully de-aerated to ensure that there is no incoming free gas. The flow parameters of the experimental campaign are listed in table \ref{Tab1}.

% ==========================Stable Ventilated Cavities ==================== %
\section{Results: Stable ventilated cavities}
The defining characteristics of stable, fixed-length ventilated cavities are discussed in this section along with their occurrence on a regime map. Transitional cavities observed during the transition from one stable type to another are discussed in section \ref{trans}.
\subsection{Cavity classification}
Four types of stable ventilated cavities are identified based on the cavity closure region for the range of $C_{qs}$ and $Fr$ considered. They are classified as foamy cavities (FC), twin-branched cavities (TBC), re-entrant jet cavities (REJC), and long cavities (LC). The classification is based on the visual interpretation of optical and X-ray snapshots corroborated by cavity characteristics, such as pressure and geometry as discussed later in subsections \ref{cavpres} and \ref{cavlength}.

\begin{figure}
    \centering    \includegraphics[width=0.70\textwidth]{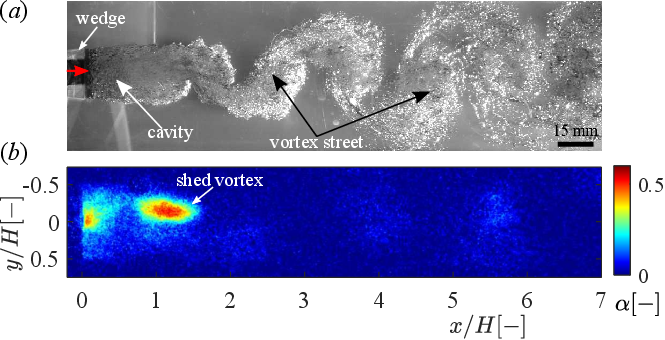}
    \caption{\label{FC_1} A side view ($x-y$)  of a foamy cavity (FC) at $Fr=13.9$ and $C_{qs}$ = 0.0205; (\textit{a}) A snapshot from high-speed optical imaging. The red arrow shows the direction of the ventilation and the bulk flow. (\textit{b}) A snapshot from high-speed X-ray imaging at the same flow condition. \textcolor{black}{The colorbar shows spanwise-averaged void fractions.}}
\end{figure}
\textit{Foamy cavities (FC)}: These cavities were observed for $C_{qs}<0.043$ and all the considered $Fr$ (2--13.9).  Figure~\ref{FC_1}(\textit{a}) shows a snapshot from high-speed imaging of a foamy cavity observed at $Fr=13.9$ and $C_{qs}$ = 0.0205. Figure~\ref{FC_1}(\textit{b}) shows the corresponding instantaneous void fraction field measured using time-resolved X-ray densitometry in a separate experiment. The densitometry is indispensable as it allows us to visualise void fractions inaccessible to conventional high-speed imaging. The cavity is characterised by the presence of injected gas as dispersed gas bubbles in the near-wake of the wedge. Foamy cavities do not have a well-defined closure region and are characterised by gas ejection via vortex shedding in the wake of the wedge. Visual observation reveals that cavities are nominally two-dimensional in the near wake region similar to natural cavities reported by \cite{Wu2021} in the same geometry. Such cavities have also been observed in other cavitator geometries like backward-facing steps \citep{Qin2019}, disc cavitators \citep{Karn2016} and three-dimensional fences \citep{Barbaca2017}.\par

\begin{figure}
    \centering    \includegraphics[width=0.73\textwidth]{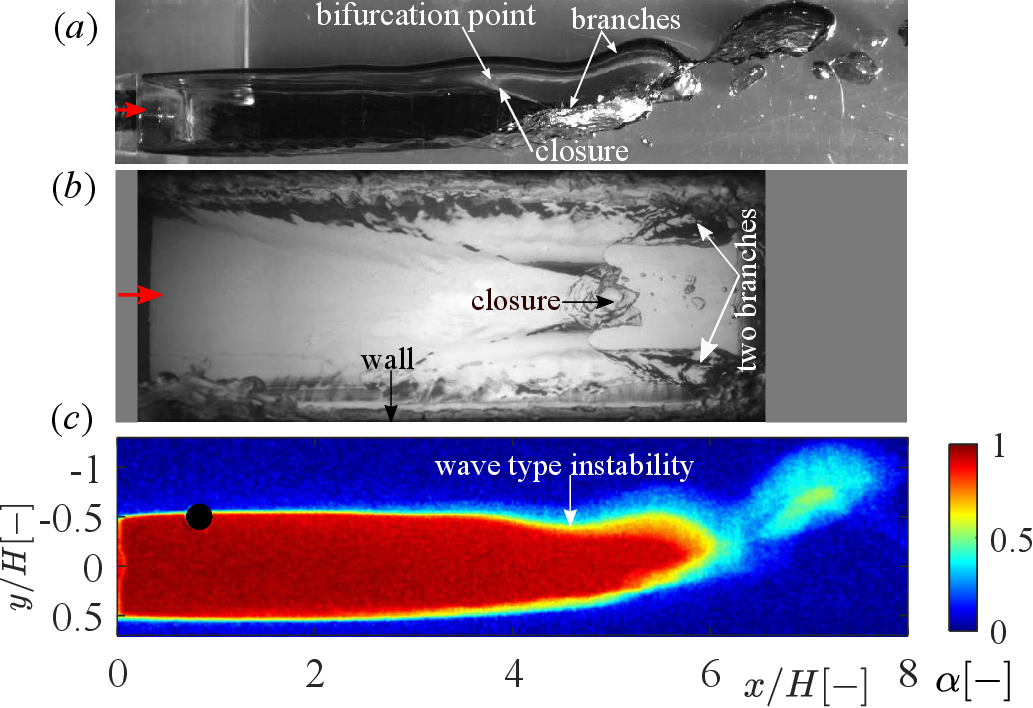}
    \caption{Twin-branched cavity (TBC) at $Fr$ = 2.08, $C_{qs}$ = 0.058; (\textit{a}) shows 
    a snapshot from high-speed optical imaging in the $x-y$ plane, (\textit{b}) shows the top view ($x-z$ plane), (\textit{c}) shows the instantaneous void fraction field of a TBC in $x-y$ plane. The black disc is a measurement artefact.}
    \label{fig:twin-cavity}
\end{figure}
\textit{Twin-branched cavities (TBC)}: %
For $2.08 \leq Fr \leq 4.17$  and $C_{qs} > 0.043$, an attached cavity with a \emph{weak} re-entrant flow near its closure (see flow structure near region marked `closure' in figure \ref{fig:twin-cavity}(\textit{b})) and two prominent branches (legs) alongside the walls are observed as shown in figure \ref{fig:twin-cavity}. A prominent travelling wave-type instability is seen on the upper cavity interface (see figure \ref{fig:twin-cavity}(\textit{c})). Twin-branched cavities are clear (filled with gas) and exhibit a prominent camber due to buoyancy effects. The resulting upward curvature of the upper cavity interface leads to lift generation and formation of trailing vortices, observed as two branches (see figure \ref{fig:twin-cavity}(\textit{b})). Similar cavities were observed behind a two-dimensional cavitator by \cite{Qin2019} and \cite{Barbaca2017}. These cavities show a close resemblance to the \textit{twin vortex tube} type ventilated cavities reported in three-dimensional axisymmetric cavitators \citep{semen, Kawakami2011, Karn2016}. With an increase in $Fr$, the upward camber of the cavity decreases due to the increased effect of fluid inertia relative to gravity. TBCs are nominally two-dimensional along their axis until the bifurcation point slightly upstream of the closure where the cavity is divided into two branches, resulting in three-dimensional cavity closure, as shown in figure \ref{fig:twin-cavity}.\par

\begin{figure}    
\centering\includegraphics[width=0.70\textwidth]{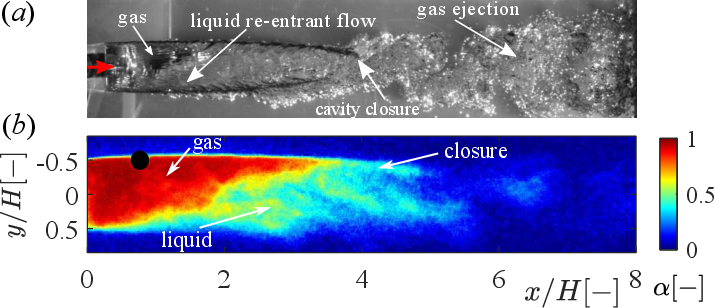}
    \caption{A side view ($x-y$) of a re-entrant jet cavity (REJC) at $Fr$ = 5.79, $C_{qs}$ = 0.054; (\textit{a}) optical imaging, (\textit{b}) an instantaneous void fraction field.}
    \label{fig:rej-cavity}
\end{figure}

\textit{Re-entrant jet cavities (REJC)}: %
A third type of cavity, characterised by a \emph{strong} re-entrant liquid flow at the cavity closure, was observed at higher flow velocity ($5.79 \leq Fr \leq 13.9$) and intermediate ventilation rate ($0.045 \leq C_{qs} \leq 0.065$). These are termed re-entrant jet cavities (REJC) and an example is shown in figure \ref{fig:rej-cavity}. The gas in the cavity can be identified by a relatively clear part (see also the labelled region in figure \ref{fig:rej-cavity}(\textit{a})), while the re-entrant flow is seen by the frothy liquid inside the cavity (see region labelled in figure \ref{fig:rej-cavity}(\textit{a})). REJ cavities are nominally two-dimensional except for the highly frothy and turbulent cavity closure. These cavities are slightly asymmetric about the wedge centerline ($y$ = 0). Despite this asymmetry, REJ cavity shapes do not have a strong dependence on $Fr$. Re-entering liquid flow is confined to the lower half of the cavity, while the gas accumulates in the upper half (see figure \ref{fig:rej-cavity}(\textit{b})). Such cavities resemble those reported by \cite{Barbaca2017} behind a three-dimensional fence. \par

\begin{figure}
    \centering
    \includegraphics[width=0.67\textwidth]{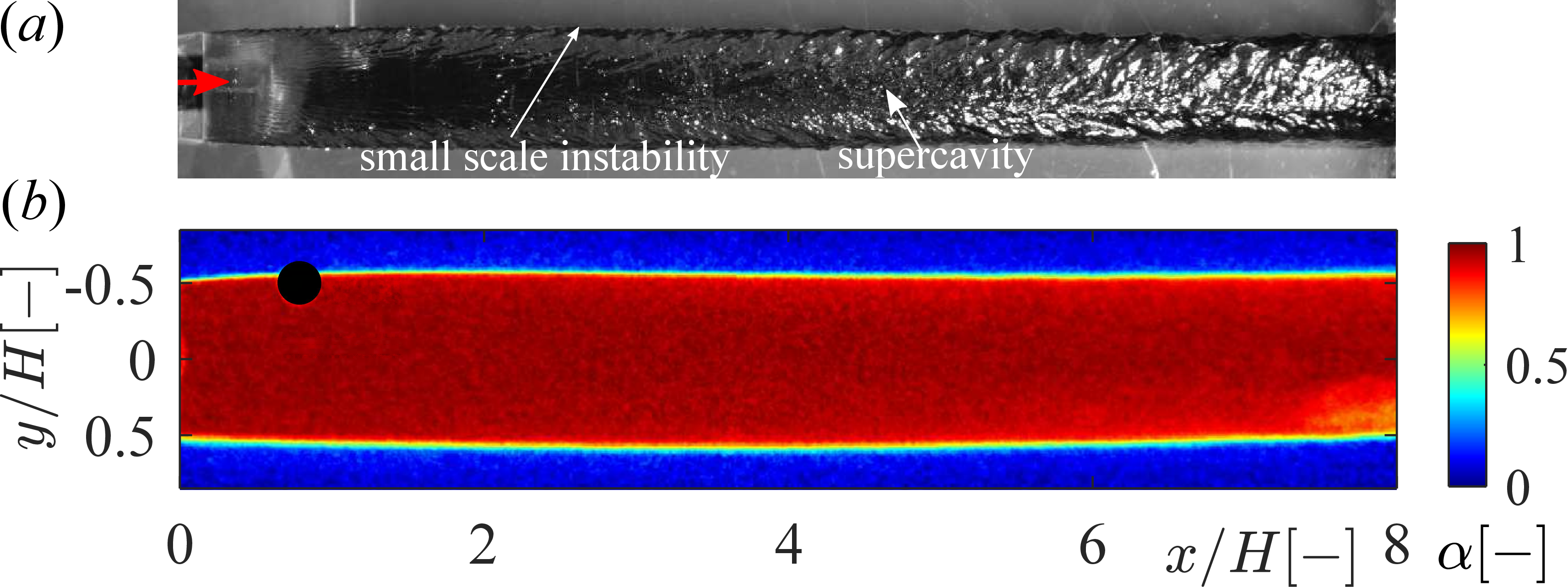}
    \caption{The side view ($x-y$) of a long cavity (LC) at $Fr$ = 10.42, $C_{qs}$ = 0.090.}
    \label{fig:long-cavity}
\end{figure}

\textit{Long cavities (LC)}: %
Long cavities exist for $5.79 \leq Fr \leq 13.9$ and $C_{qs}\geq$ 0.07. LC span beyond the optical FOV (see figure \ref{fig:long-cavity}). Thus, the complete cavity could not be visualised with high-speed imaging and cavity closure could not be measured quantitatively using X-ray densitometry. However, qualitative X-ray-based visualisation was performed to study cavity closure dynamics. Typically, the length of such cavities is more than 12$H$. The observable portion of these cavities was two-dimensional, filled with gas as evident from the instantaneous void fraction distributions. There is no observable effect of gravity on their shape. These cavities have small-scale instability/waves on the cavity interface, as indicated in figure \ref{fig:long-cavity}(\textit{a}). Long cavities exhibit oscillation in the $x-y$ plane similar to pulsating/vibrating cavities reported by \cite{silbersong61}, \cite{Michel1984}, and \cite{skid}. 

\subsection{Cavity \textcolor{black}{closure} regime map}
\begin{figure}
    \centering
    \includegraphics[width=0.57\textwidth]{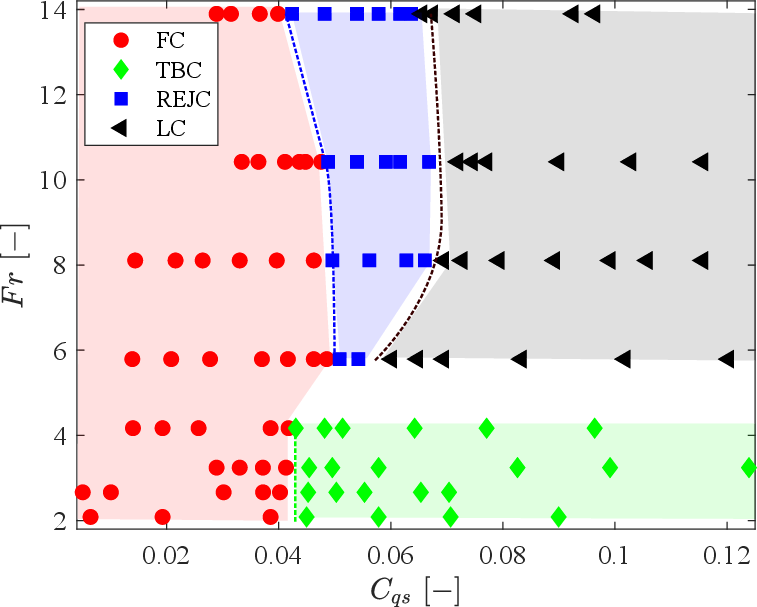}
    \caption{The regime map ($Fr$--$C_{qs}$) shows different types of observed ventilated partial/super--cavities indicated by different colors; red: foamy cavities (FC), green: twin-branched cavities (TBC), blue: re-entrant jet cavities (REJC), black: long cavities (LC). The dashed lines show the demarcation from one regime to another, i.e. green dashed line: $C^{tr}_{qs,fc-tbc}$, blue dashed line: $C^{tr}_{qs, fc-rejc}$, black dashed line: $C^{tr}_{qs, rejc-lc}$ }
    \label{fig:regime} 
\end{figure}
The observed cavity \textcolor{black}{closure} types and transition regions in between them are identified on a regime map defined by the Froude number ($Fr$) and ventilation coefficient ($C_{qs}$) as shown in figure ~\ref{fig:regime}. Note that this regime map is specific to the cavitator geometry under consideration, \textcolor{black}{i.e. wedges.} The regime map was generated by fixing $U_{0}$ ($Fr$), followed by increasing the gas injection to achieve a $C_{qs}$ following the L-H ventilation strategy explained in subsection \ref{procedure} (red profile in figure~\ref{fig:injection-profile}). Note that every data point is an independent experiment, i.e. with a ventilation profile starting from $C_{qs}$ = 0. For $Fr \gtrsim 5$, the effect of gas buoyancy was observed to be less pronounced, and thus cavity types observed for $Fr \lesssim 5$ and $Fr \gtrsim 5$ were different. 
FCs (red region) and TBCs (green region) were observed for $Fr \lesssim 5$, meaning TBC is the only supercavity observed at low $Fr$. FC (red region), REJC (blue region), and LC (grey region) were observed for $Fr \gtrsim 5$. Thus two types of supercavities were observed at high $Fr$, namely  REJC and LC. The transition between the regimes occurs at a critical Froude number of $\sim$ 5 and near the dashed lines shown in figure~\ref{fig:regime}. \par

\subsection{Cavity pressure} \label{cavpres}
\begin{figure*}
    \centering
    \includegraphics[width=0.99\textwidth]{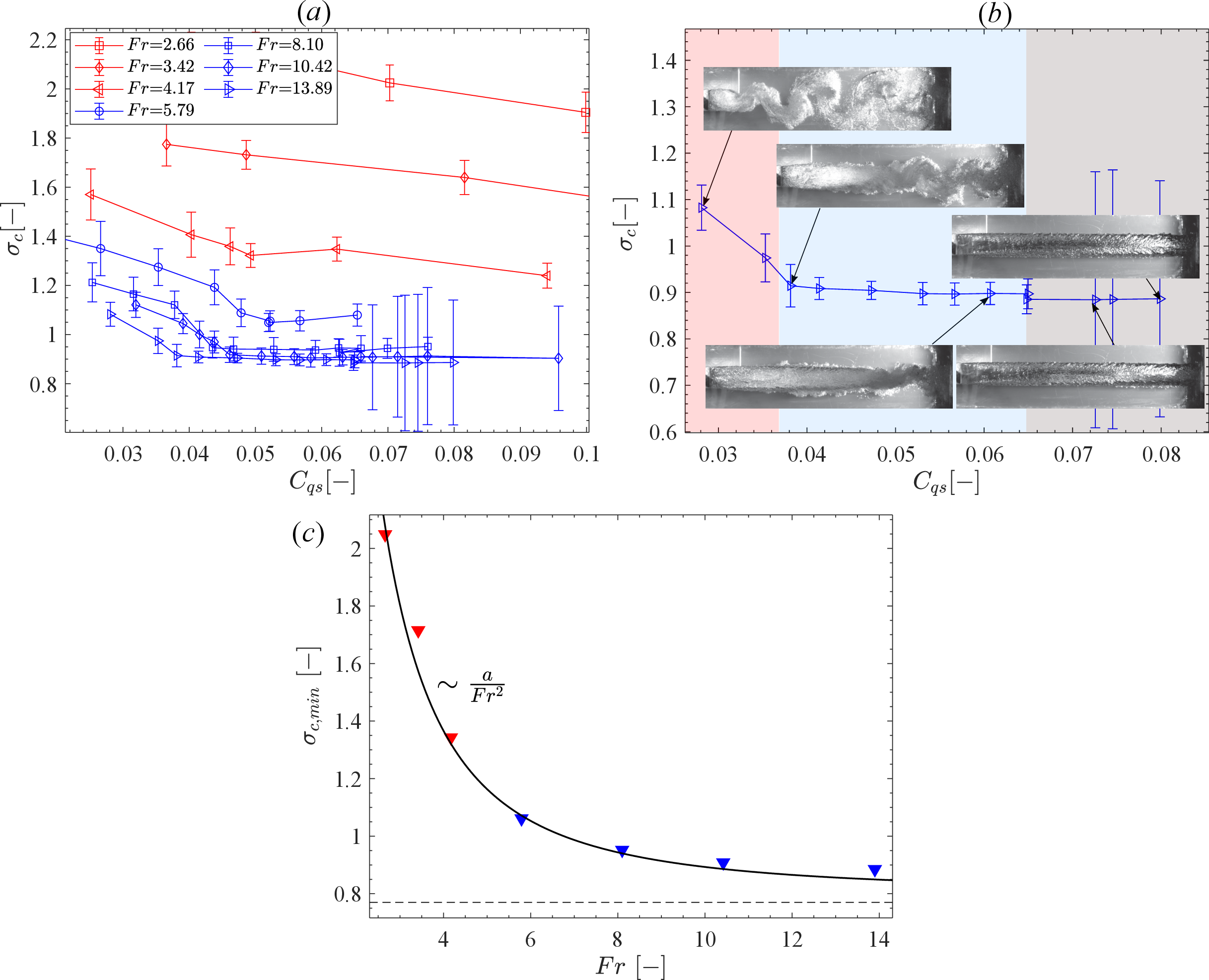}
    \caption{(\textit{a}) Cavitation number ($\sigma_{c}$) based on measured cavity pressure as a function of ventilation coefficient ($C_{qs}$) at different $Fr$ (red markers show low $Fr$ cases, while blue markers show high $Fr$ cases). (\textit{b}) $\sigma_{c}$ at $Fr = 13.89$ showing different ventilated cavity regimes. The red, blue, and grey regions correspond to FC, REJC and LC closures respectively, (\textit{c})  minimum $\sigma_{c}$ as a function of $Fr$. The black curve shows a fit of the form $aFr^{-2}$. The black dashed line shows $\sigma_{c}$=0.77 computed with the Bernoulli equation.}
    \label{fig5}
\end{figure*}

The pressure in the ventilated cavity is measured on the wedge centerline near the base, marked $P_{c}$ in figure~\ref{fig:setup}(\textit{a}). The measured pressures (\textcolor{black}{$P_{c}$ and $P_{0}$}) are used to compute the cavitation number ($\sigma_{c}$). 
Figure~\ref{fig5}(\textit{a}) shows the variation of $\sigma_{c}$ with $C_{qs}$ for a range of $Fr$ considered in this study: the low $Fr$ cases are shown by red markers, while the high $Fr$ cases are shown by blue markers. This demarcation in low versus high $Fr$ reflects the observations in the regime map (figure \ref{fig:regime}), which show a change in cavity closure type near $Fr$ = 5. 
At low $Fr$ ($\approx 2.08-4.17$), the cavity pressure could only be measured for TBCs, as foamy cavities at low $Fr$ were short, barring reliable measurement at the $P_{c}$ location. Cavity pressure is seen to be the highest (high $\sigma_c$) at the lowest $Fr$. 

For FC, the cavity pressure oscillates about a mean due to the periodic vortex street, however, for TBC and REJC the cavity pressure remains fairly constant. 
For LC, pressure oscillations are the largest due to the cavity pulsation in the $x-y$ plane. 
These pulsations are evident from the large standard deviations in $\sigma_{c}$ for long cavities as shown in figure~\ref{fig5}(\textit{a,b}). 
At lower $Fr$, once TBC is formed, $\sigma_{c}$ does not change significantly for increasing $C_{qs}$ as shown in figure~\ref{fig5}(\textit{a}).
Figure~\ref{fig5}(\textit{b}) shows the variation of $\sigma_{c}$ with $C_{qs}$ along with flow visualisations for a fixed Froude number ($Fr = 13.89$): $\sigma_{c}$ decreases with an increase in ventilation ($C_{qs}$) in the FC regime. With a further increase in $C_{qs}$, $\sigma_{c}$ decreases minimally in the REJC regime until it reaches the asymptotic minimum ($\sigma_{c, min}$) when the formation of LC occurs. 
In the LC regime, $\sigma_{c}$ stays constant at $\sigma_{c, min}$. Figure~\ref{fig5}(\textit{c}) shows the minimum cavitation number, $\sigma_{c, min}$, attained at a given $Fr$. For LC (high $Fr$ and $C_{qs}$), the streamlines near the wedge base tend to straighten towards being parallel to the incoming flow (see figure \ref{fig:long-cavity}(\textit{b})) and the pressure in the cavity approaches a minimum. 
With streamlines parallel to the tunnel walls, the liquid pressure outside of the air-cavity, $P_\ell$, can be estimated from the Bernoulli equation using the wedge's solid blockage percentage, $\xi$.
The pressure is given as $P_\ell = P_0 + 1/2\rho_\ell U_0^2\left[1 -\left(1-\xi\right)^{-2} \right]$.
Here, $\xi = 0.25$ leads to ($P_0-P_\ell)/(1/2\rho U_0^2)\approx 0.77$.
This value is indicated by the horizontal black dashed line in figure~\ref{fig5}(\textit{c}).
Thus, as the cavity interface becomes \emph{straighter}, $P_c$ approaches the theoretical $P_\ell$. 

\subsection{Cavity geometry} \label{cavlength}
The void fraction fields (instantaneous and time-averaged) of the cavity along with high-speed visualisations, are used to study various geometrical aspects of the cavity, such as length, thickness, and shape. The lower and higher $Fr$ cases are discussed separately for clarity. \par
\subsubsection{Cavity length} 
\begin{figure*} 
    \centering
    \includegraphics[width=0.94\textwidth]{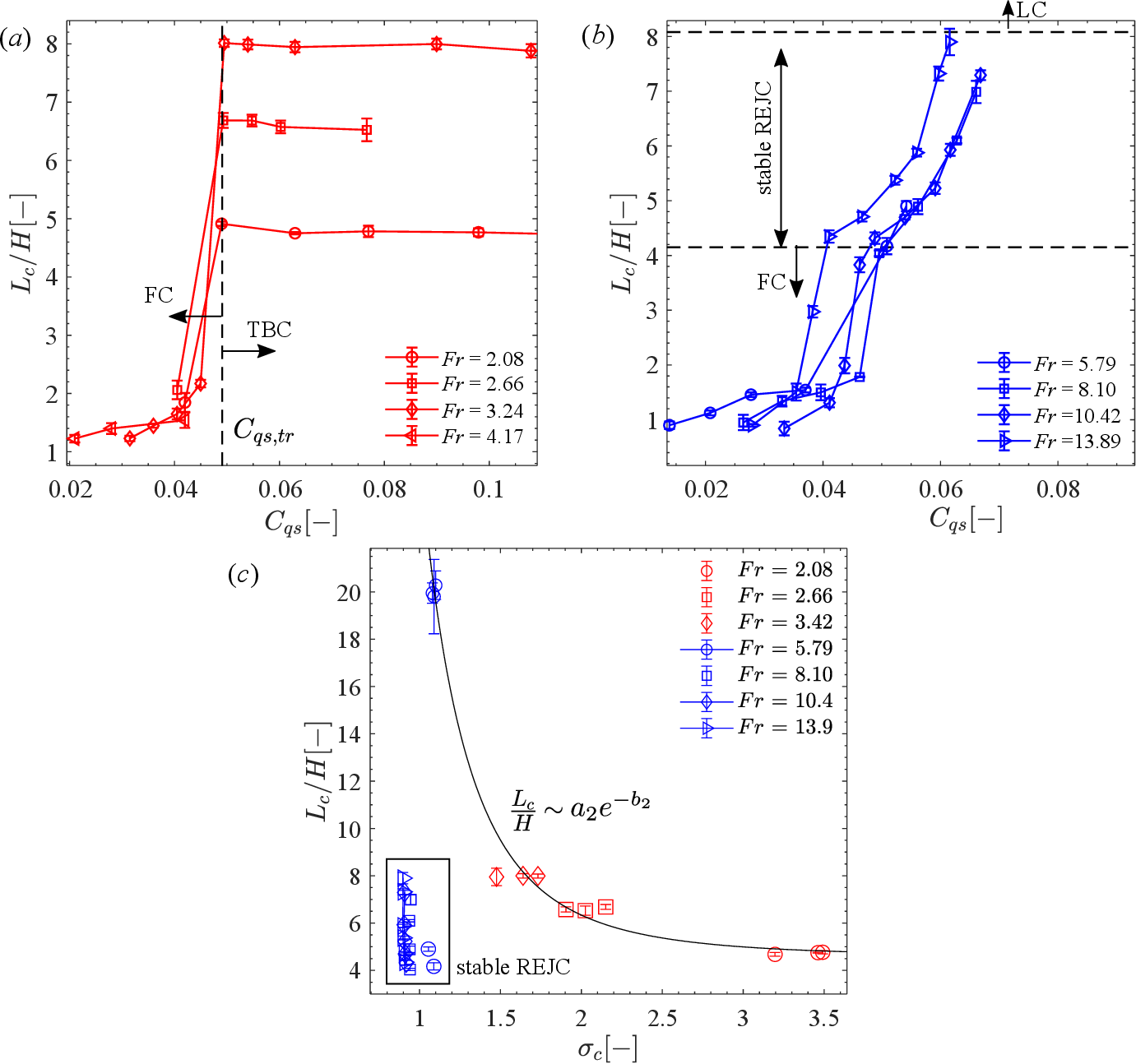}
    \caption{The length ($L_{c}$) of the partial ventilated cavities, normalised by wedge height ($H$) at different $Fr$ as a function of ventilation coefficient ($C_{qs}$): (\textit{a}) shows cavities at low $Fr$ ($<$ 5), while (\textit{b}) shows cavities at higher $Fr$ ($>$ 5). (\textit{c}) Supercavity length variation with cavitation number ($\sigma_{c}$) (the low $Fr$ cases are shown by red markers, while the high $Fr$ cases are shown by blue markers). \textcolor{black}{The black curve shows a power law fit.}} 
    \label{fig:cavity-length}
\end{figure*}

The cavity length is defined by the distance of the two-dimensional closure region from the wedge base for FC, REJC and LC. However, for TBC, cavity length is defined from the wedge base to the bifurcation point along the mid span of the cavity. Non-dimensional cavity length ($L_{c}/H$) variation with $C_{qs}$ for $Fr \leq 4.17$ and $Fr \geq 5.8$ is shown in figure~\ref{fig:cavity-length}(\textit{a}) and (\textit{b}), respectively. Figure~\ref{fig:cavity-length}(\textit{c}) shows the variation of ($L_{c}/H$) with cavitation number ($\sigma_{c}$) for supercavities.\par
For $Fr \leq 4.17$, the length of the foamy cavities increase monotonically with $C_{qs}$. The absence of a well-defined closure leads to a modest increase in $L_{c}/H$ with increasing $C_{qs}$. $L_{c}/H$ increases abruptly when FC transitions to TBC, and with a further increase in $C_{qs}$ at a fixed $Fr$, $L_{c}/H$ remains unchanged (see figure \ref{fig:cavity-length}(\textit{a})). The observed trend is accompanied by the thickening of the two branches and a higher gas ejection rate needed to maintain the same cavity length. 
At  $Fr \geq 5.8$, the length of foamy cavities increases monotonically with $C_{qs}$ similar to the low $Fr$ case (figure \ref{fig:cavity-length}(\textit{b})). 
A sharp change in cavity length, from $L_{c}/H \approx$ 2 to 4, with an increase in $C_{qs}$ is observed as the cavity regime changes from FC to REJC. As $C_{qs}$ is increased further, REJ cavities (figure \ref{fig:cavity-length}(\textit{b})) increase in length, and are seen to span over $4H$ to $8H$ for all $Fr$ considered. 
Note that the length of the cavity is the same as the length of the re-entrant jet for REJ cavities.
With further increase in $C_{qs}$ ($\gtrsim$ 0.07), abrupt growth of REJ cavities results in the formation of long cavities that grow out of the field of view. 
It was not possible to quantitatively measure the length of LCs due to limited optical and X-ray densitometry access. 
However, qualitative visualisation of closure using X-ray imaging allowed the estimation of cavity length for long cavities for $Fr \approx$ 5.79.  \par
The supercavity (TBCs and LCs) lengths attains a maximum for a given $Fr$ consistent with the observation of \cite{Qin2019} in two-dimensional cavitator.
Furthermore, the measured maximum cavity length (for a fixed $Fr$) appears to have a power law relationship with the cavitation number, $\sigma_{c}$, as shown in figure~\ref{fig:cavity-length}(\textit{c}) and reported previously by \cite{Terentiev2011}.
Interestingly, REJ cavities appear \emph{stunted} as they do not fall on the power law, suggesting a different scaling law governing their lengths.
This will be addressed in subsection \ref{lcform}. 
The maximum {\em{thickness}} of cavities shows a minor dependence on $C_{qs}$: it varies from $1-1.1 H$ for FC, while it is close to $1.18H$ for all other supercavities observed in this study. This is in agreement with \cite{Wu2021} who reported the geometric features of natural cavities in the same geometry.\par

\subsubsection{Average void fraction distribution}
\begin{figure}
    \centering
    \includegraphics[width=1.00\textwidth]{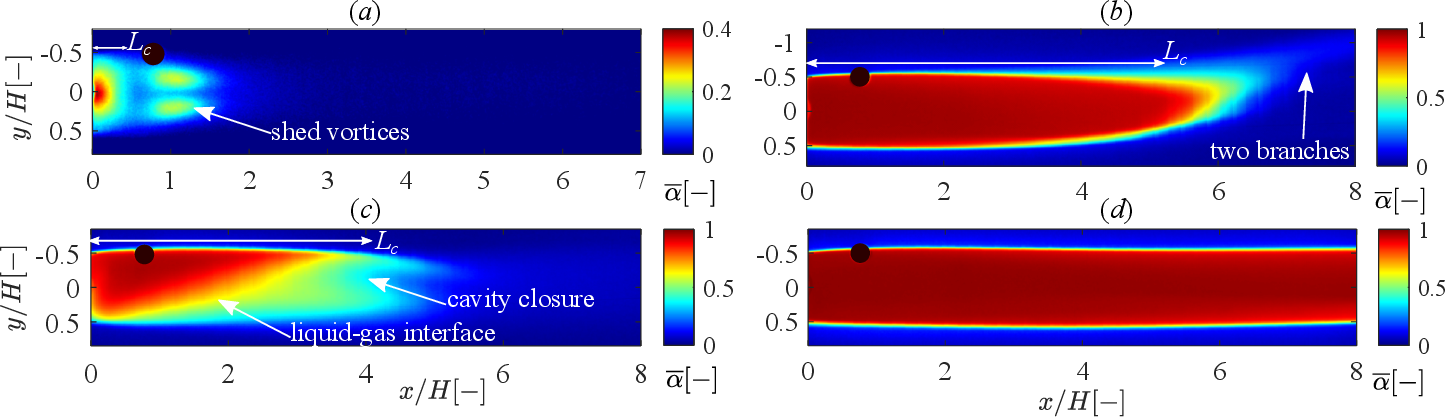}
    \caption{Time-averaged void fraction fields of cavities in the four regimes; (\textit{a}) FC: $Fr$ = 13.9, $C_{qs}$ = 0.0205, (\textit{b}) TBC: $Fr$ = 2.08, $C_{qs}$ = 0.058, (\textit{c}) REJC: $Fr$ = 5.79, $C_{qs}$ = 0.054, (\textit{d}) LC: $Fr$ = 10.42, $C_{qs}$ = 0.090. The black disc is a measurement artefact. Note that panel ($a$) uses different limits of the colorbar.} 
    \label{avg_vc}
\end{figure}
Figure \ref{avg_vc} shows the time-averaged and spanwise integrated void fraction, $\overline{\alpha}$, for the ventilated cavities under consideration. The black disk located at about ($x/H$=0.78, $y/H$= -0.52) in the void fraction fields is a measurement artefact as mentioned before. For FCs, the maximum of time-averaged void fractions is $\sim$ 0.4 as cavities are composed of dispersed bubbles in the near wake, as shown in figure \ref{avg_vc}(\textit{a}).
% Lower than what?
The gas-filled vortex streets appear as lobes with an average void fraction of $\sim$ 0.2 and the time-averaged void fraction is symmetric about the $x$-axis for foamy cavities. 
TBC, REJC, and LC are composed predominantly of air, resulting in  $\alpha$ = 1 as shown in figure \ref{avg_vc}(\textit{b}-\textit{d}). For twin-branched cavities (see figure \ref{avg_vc}(\textit{b})), the void fraction distribution is asymmetric about the $x$-axis due to the upward camber as a result of buoyancy.  
$\overline{\alpha}$ in the closure region ranges from approximately $0.1-0.6$ and this variation is due to the averaging of the three-dimensional, time-varying closure geometry as shown in figure \ref{fig:twin-cavity}(\textit{b}). 
Figure~\ref{avg_vc}(\textit{c}) shows the void fraction distribution of REJ cavities. There is a clear separation of gas and liquid phases inside the cavity due to the buoyancy effects as seen by a sharp gas-liquid interface. The gas ejected via spanwise vortices near the closure appears as lobes, asymmetric about $y=0$. 
The cavity shape exhibits asymmetry about $y=0$, with a downward curvature of the upper cavity interface. 
The downward curvature of the cavity interface relaxes as cavity length increases to form a long cavity. The streamlines of long cavities appear relatively straighter, with the shape of the cavity almost symmetric about $y=0$, as shown in figure \ref{avg_vc}(\textit{d}). The void fractions at the closure of long cavities could not be measured due to the limitations imposed by the FOV. Nevertheless, qualitative gas distribution is discussed in subsection \ref{closure}.\par

\subsection{Dynamics at cavity closure and gas ejection mechanisms}
\begin{figure*}
    \centering   \includegraphics[width=0.99\textwidth]{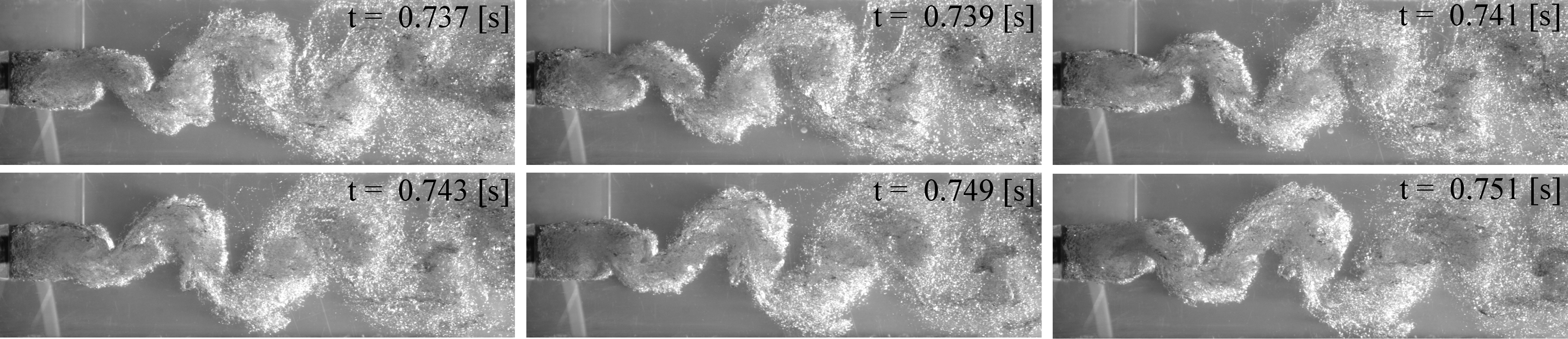}
    \caption{High-speed flow visualisation of gas ejection for FC at $Fr$ = 13.9, $C_{qs}$  = 0.0290, showing periodic gas ejection via a Von K\'arm\'an vortex street.}
    \label{fig_fc_hs}
%\end{figure*}
%\begin{figure*}
    \vspace{3mm}
    \centering \includegraphics[width=0.96\textwidth]{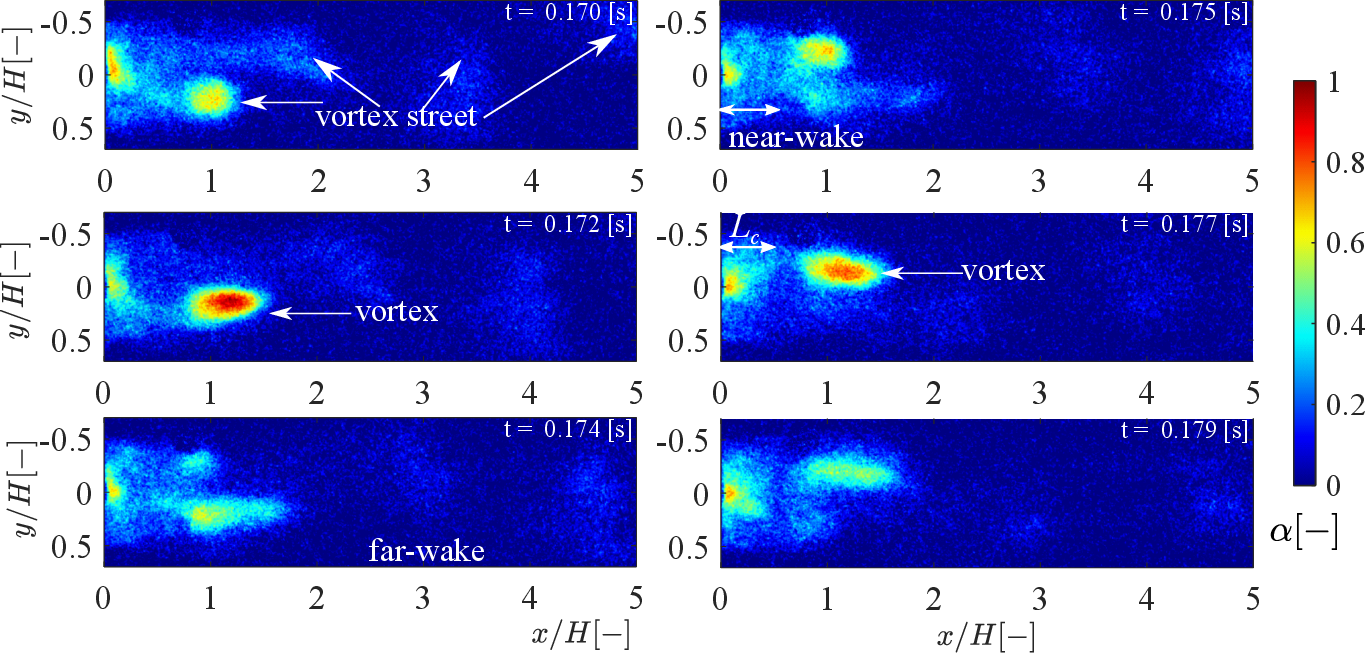}
    \caption{Instantaneous void fraction fields for FC at $Fr$ = 13.9, $C_{qs}$  = 0.0205, showing periodic gas ejection via a Von Kármán vortex street.}
    \label{fig_fc}
\end{figure*}

\textit{Foamy cavities}: 
Foamy cavities do not have a well-defined closure, akin to natural \textit{open} cavities defined by \cite{Laberteaux2001a}. \textcolor{black}{These are wake-dominated cavities wherein, the entrained gas is ejected out periodically via an alternating Von K\'arm\'an vortex street.} This is shown by high-speed snapshots in figure~\ref{fig_fc_hs} (supplementary movie S1) and void fraction time-series in figure~\ref{fig_fc} (acquired in a separate set of experiments, see supplementary movie S2). The gas ejection frequency, expressed as Strouhal number ($St_{H}$), is estimated to be $\sim 0.31$, which is in close agreement with values reported for other bluff bodies by \cite{Brandao2019, Wu2021}. 

\begin{figure}
    \centering\includegraphics[width=0.85\textwidth]{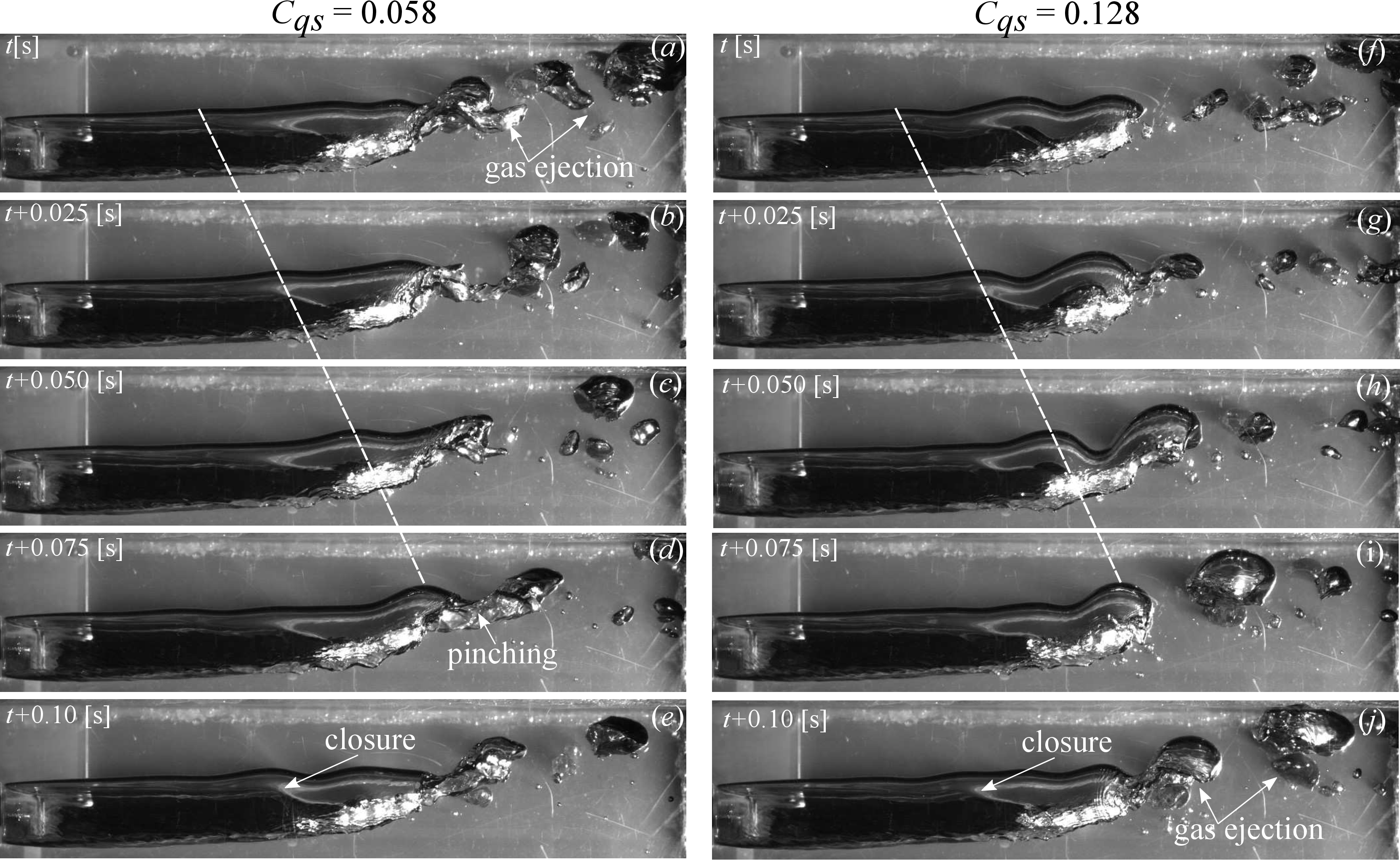}
    \caption{High-speed visualisation of TBC closure at $Fr$ = 2.08 for two different cases of ventilation; left (\textit{a-e}): $C_{qs}$ = 0.058, right (\textit{f-j}) : $C_{qs}$ = 0.128. The white dashed line shows a convecting wave-type instability on the upper cavity interface responsible for cavity pinching.}
    \label{tbc_comp}
    \vspace{3mm}
%\end{figure}
%\begin{figure}
    \centering
    \includegraphics[width=0.97\textwidth]{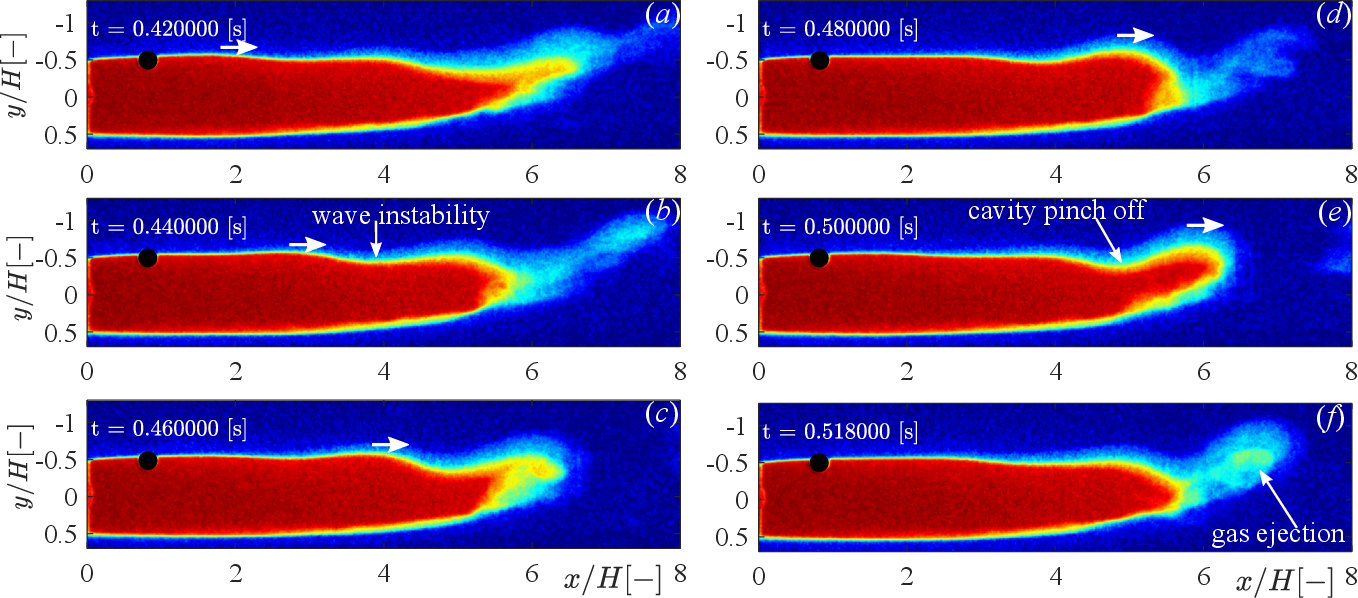}
    \caption{Instantaneous gas ejection in TBC closure at $Fr$ = 2.08, $C_{qs}$  = 0.058. The white arrows show the traveling wavefront. The colorbar of the jet colormap span from 0 to 1.}
    \label{tbc_dynamic}
\end{figure}

\textit{Twin-branched cavities (TBC)}: % Temporarily added subsubsections to show in outline.
Twin-branched cavities are nominally two-dimensional until their bifurcation point but have a well-defined three-dimensional closure. The closure of these cavities is characterised by two branches along the walls as shown in figure \ref{fig:twin-cavity}  and \ref{tbc_comp}. The cavity closure is marked by a weak re-entrant flow along the centreline of the cavity in the $x-z$ plane shown in figure \ref{fig:twin-cavity}(\textit{b}).
Based on high-speed videos, two gas ejection mechanisms can be identified. Firstly, the gas ejection appears to be due to a re-entrant flow in the narrow mid-tail region, where gas is ejected in small parcels as shown in the region marked by `closure' in figure~\ref{fig:twin-cavity}(\textit{a, b}). See supplementary movie S3 for the visualisation of cavity span ($x-z$ plane).
Secondly, gas ejection is observed at the downstream tips of the two branches via a wave-like instability shown in figure \ref{tbc_comp} and \ref{tbc_dynamic}. Figure~\ref{tbc_dynamic}(\textit{a})-(\textit{f}) and supplementary movie S4 shows X-ray-based visualisation of the wave-induced gas ejection in twin-branched cavities. The convecting wavefront on the upper cavity interface is marked and shown by white arrows. As the travelling wave disturbance propagates downstream along the cavity (\textit{a})-(\textit{e}), it pinches the cavity at the twin branches leading to a gas ejection. Since the length of the cavity does not change with increasing $C_{qs}$ for a given $Fr$ (see figure~\ref{fig:cavity-length}(\textit{a})), larger gas ejection must occur at higher $C_{qs}$. The larger gas ejection at $C_{qs}=0.128$ appears as larger slugs as shown in figure~\ref{tbc_comp}(\textit{f-j}). It is observed that the amplitude of the instability is higher for larger ventilation ($C_{qs}$) for a given $Fr$, suggesting a higher degree of pulsation at higher $C_{qs}$. \textcolor{black}{This is seen clearly by comparing the closure dynamics in the supplementary movies S5 and S6.} The convecting wavefront is marked by a white dashed line in figure \ref{tbc_comp} for $C_{qs}$ = 0.058 (\textit{a--e}) and 0.128 (\textit{f--j}). The slope of the white dashed lines can be used to approximate the propagation speed of the wave-instability and appears to be near identical, despite a significant difference in $C_{qs}$. For a given $Fr$, the wave propagation velocity is close to the velocity of the upper cavity interface ($\mathcal{O}(U_{0})$) for the considered range of $C_{qs}$ ($\approx$ 0.058--0.128). This suggests that it is an inertial wave possibly triggered by a Kelvin-Helmholtz instability due to the density difference at the upper cavity interface.
The propagation speed of instability ($c_{r, tbc}$) at the interface of two fluids in the absence of a bounding wall is given by \ref{cr_tbc} \citep{Drazin2002}, where subscripts $l$, $g$ correspond to liquid and gas phases respectively. 
\begin{equation}  \label{cr_tbc}
    c_{r, tbc} = \frac{\rho_{g}U_{g} + \rho_{l}U_{l}}{\rho_{g}+ \rho_{l}} \approx U_{l} \approx U_{0}
\end{equation}
For $\rho_{g} \ll \rho_{l}$, $c_{r, tbc}$ reduces to $ U_{l}$ due to the large density difference across the cavity interface. This is consistent with our experimental observations, i.e. propagation velocity of instability is of the order of inflow velocity.\par

\begin{figure*}  
    \centering
    \includegraphics[width=0.95\textwidth]{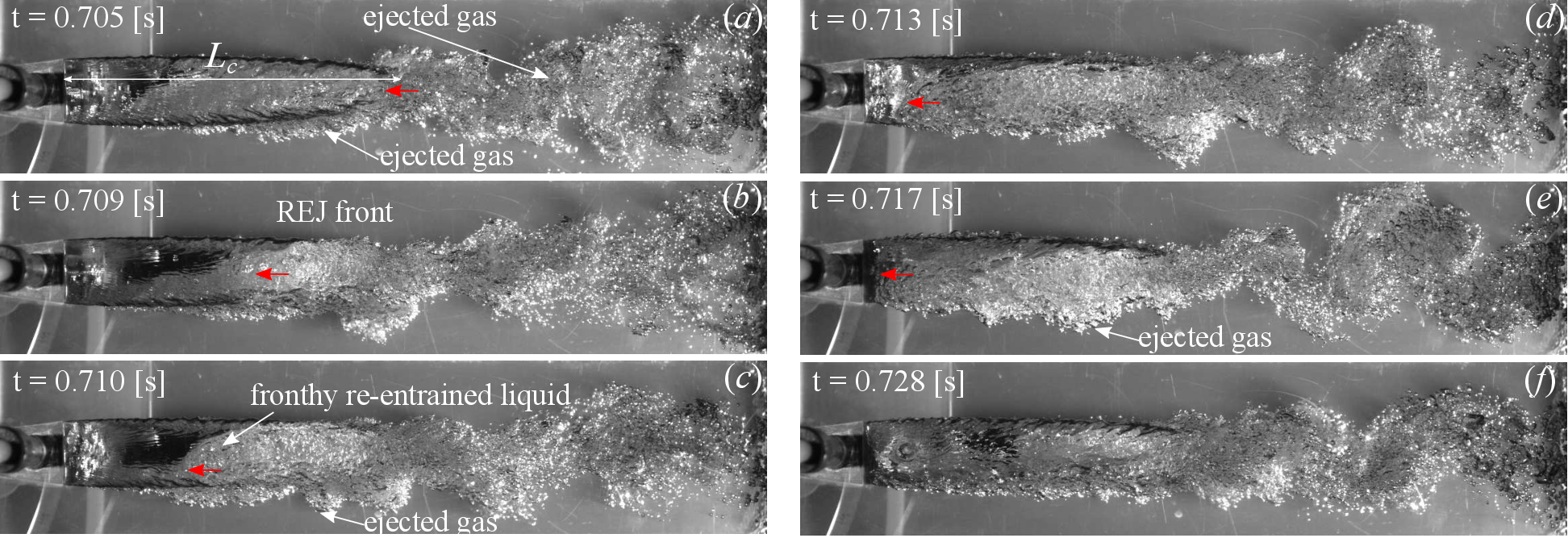}
    \caption{High-speed visualisation of REJC dynamics at $Fr = 5.79$, $C_{qs} = 0.054$ (time increases from top to bottom, left to right). The red arrows in (\textit{a-d}) indicate the re-entrant flow. The Von K\'arm\'an vortex street can be clearly seen in the wake of cavity.}
    \label{fig:rej-timeseries}
%\end{figure*}
 \vspace{3mm}
%\begin{figure*}   % see 13/01/2023, F23
    \centering
    \includegraphics[width=0.98\textwidth]{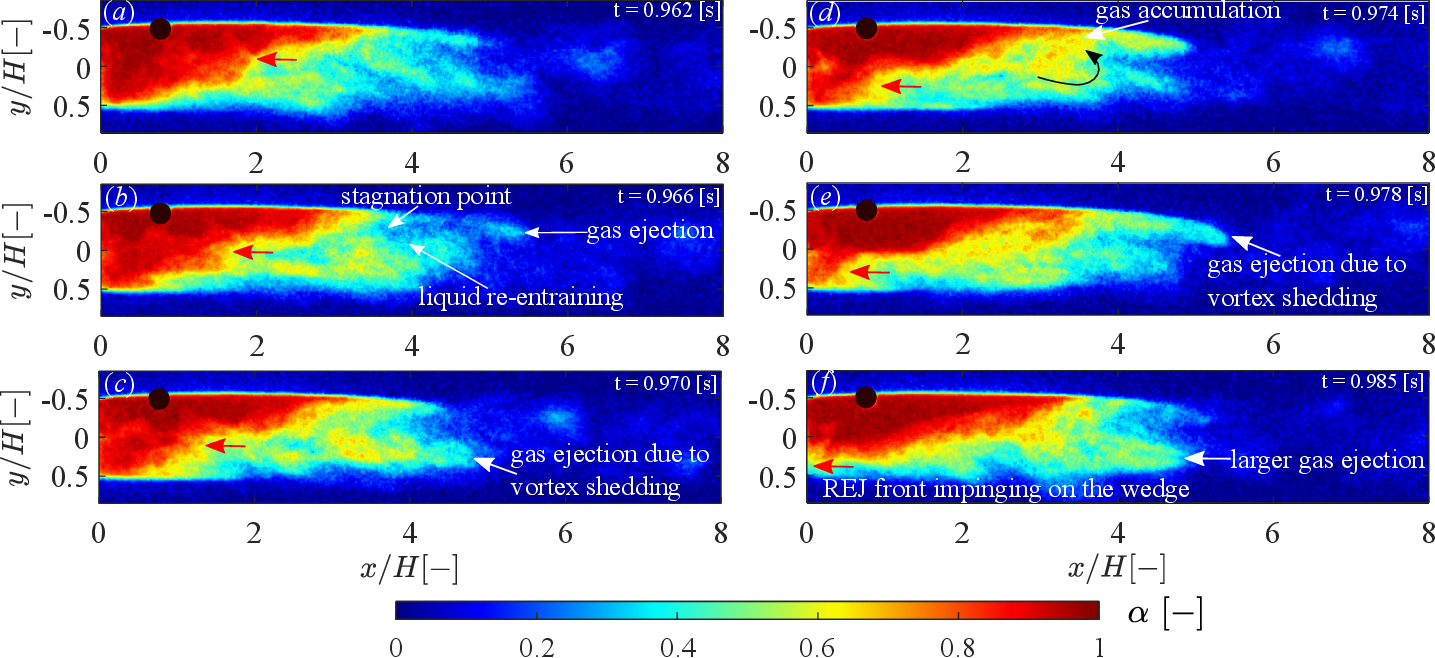}
    \caption{Instantaneous void fraction fields for a REJ cavity at $Fr$ = 5.79, $C_{qs}$ = 0.054 showing gas accumulation and ejection. Red arrows indicate the re-entrant flow front propagating upstream.}
    \label{fig:rej-timeseries-xray}
\end{figure*}

\begin{figure*} % see 13/01/2023, F23
    \centering
    \includegraphics[width=0.97\textwidth]{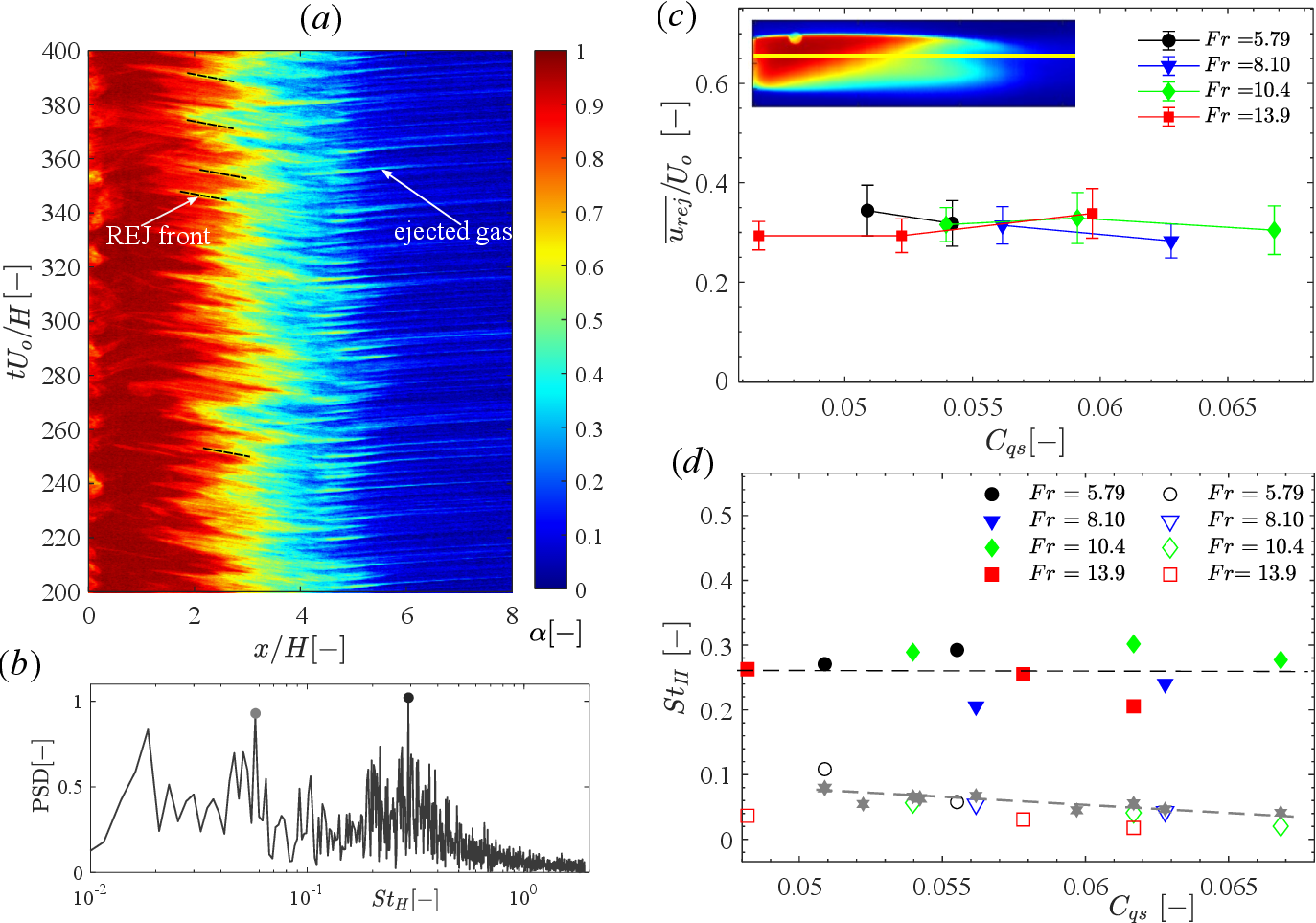}
     \caption{(\textit{a}) A sample $x-t$ diagram showing the evolution of the void fraction for a REJ cavity at $Fr = 5.79$, $C_{qs} = 0.054$. Examples of the re-entrant flow front are indicated by black dashed lines. (\textit{b}) Power spectral density of the void fraction computed from the $x-t$ plot at $X/H \sim$ 5, showing two \emph{distinct} 
$St_{H}$ peaks. (\textit{c}) Re-entrant flow velocity estimated from the $x-t$ diagram. (\textit{d}) Non-dimensional shedding frequency ($St_{H}$) for a range of $Fr$ and $C_{qs}$. Solid markers show the gas ejection frequency due to vortex shedding. Open markers show the gas ejection frequency due to the re-entrant jet impingement. Grey markers ($\ast$) show $St$ based on measured kinematics of the re-entrant liquid flow ($L_{c}$ and $\overline{u_{rej}}$). The grey dashed line fit to grey markers is given as $St_{H,rej} = 0.2-2.375C_{qs, in}$.}
    \label{fig:rej-spacetime}  
\end{figure*}

\textit{Re-entrant jet cavities (REJC)}: % 
A time-series obtained from high-speed videography of a REJ cavity is shown in figure~\ref{fig:rej-timeseries}. 
In the optical measurements, gas-liquid interfaces result in reflections (bright, white regions) which obfuscate the internal cavity flow. However, the large field of view allows for the wedge base, cavity, and wake to be simultaneously visualised. A re-entrant liquid flow, visualised by a frothy mixture along with the flow-front indicated by a red arrow can be seen propagating upstream in figure~\ref{fig:rej-timeseries}(\textit{a}-\textit{e}). As the re-entrant liquid flow reaches the base of the wedge it seems to be redirected to the cavity interfaces as sshown in figure~\ref{fig:rej-timeseries}(\textit{e}). This is illustrated in the supplementary movie S7. A clearly defined Von K\'arm\'an vortex street can be seen in the cavity wake (figure~\ref{fig:rej-timeseries}(\textit{d},\textit{e})), suggesting (i) vortex shedding and gas ejection are coupled, \textcolor{black}{(ii) strong wake-gas interaction in such cavities.} As the re-entrant jet front propagates upstream, it displaces gas at the bottom of the cavity as seen in figure \ref{fig:rej-timeseries}(\textit{c,e}). While the gas can be seen advecting away from the cavity, the gas volume cannot be quantified from the high-speed imaging.

Time-resolved void fraction measurements of a REJ cavity for the same condition show the internal cavity flow structures in figure~\ref{fig:rej-timeseries-xray} and supplementary movie S8. Note that the X-ray densitometry measurements are separate experimental trials from the high-speed videography measurements. The gas accumulation in the upper half of the cavity is evident from the red (high-void fraction) regions in figure~\ref{fig:rej-timeseries-xray}. Supplementary movie S8 shows that the injected gas reaches the closure and gets accumulated in the upper part of the cavity, as also indicated by figure~\ref{fig:rej-timeseries-xray}(\textit{d}). The liquid re-entrant flow is visualised by a low void fraction ($\sim$ 0.5) region in the lower part of the cavity. A stagnation point is formed near the closure (indicated in figure~\ref{fig:rej-timeseries-xray}(\textit{b})) and liquid re-entrant flow is pushed inside the cavity. The front propagates towards the wedge base and eventually becomes confined to the bottom part of the cavity due to the effects of gravity (see figure~\ref{fig:rej-timeseries-xray}). Here the red arrows indicate the liquid re-entrant flow front. Figure~\ref{fig:rej-timeseries-xray}(\textit{f}) shows that the propagating re-entrant liquid flow displaces the gas, resulting in a gas ejection. Additionally, gas is also ejected from the cavity via the spanwise vortices of the wake flow downstream as shown in figure~\ref{fig:rej-timeseries-xray}(\textit{b},~\textit{e},~\textit{f}). This re-entrant flow cycle continues to sustain a stable, \textcolor{black}{fixed-length} REJ cavity.

The characteristics of the liquid re-entrant flow can be estimated from a space-time ($x-t$) diagram of time-resolved void fraction fields. 
The $x-t$ plot in figure~\ref{fig:rej-spacetime}(\textit{a}) shows the void fraction time evolution along the yellow line ($y/H \approx$ - 0.27) shown in the inset of figure~\ref{fig:rej-spacetime}(\textit{c}).
% It's averaged along the box y-extent, but how? 
The black dotted lines visualise the liquid REJ fronts \emph{inside} the cavity, and the ejected gas is indicated by the white arrow in the $x-t$ plot. 
The velocity of the re-entrant jet front and the convection of the ejected gas are estimated from such $x-t$ diagrams. 
The measured REJ front velocity $u_{rej}$, when normalised with the inflow velocity ($U_{0}$), is constant ($\overline{u_{rej}}/U_0 \approx 0.3$) irrespective of $C_{qs}$ and $Fr$ as shown in figure~\ref{fig:rej-spacetime}(\textit{c}). Strouhal number ($St_{H} = fH/U_0$) of the gas ejection frequency, $f$, estimated via an FFT of the time signal extracted from the $x-t$ plot, is shown in figure~\ref{fig:rej-spacetime}(\textit{d}). The FFT was performed on the time signal of the void fraction at a probing point in the wake ($X/H \approx$ 5). Two distinct peaks were observed in the spectral density (see figure \ref{fig:rej-spacetime}(\textit{b})) corresponding to two distinct gas ejection mechanisms.

Gas ejection due to vortex shedding occurs at a roughly constant non-dimensional frequency ($St_{H,v}\approx 0.27$). The gas ejection associated with the re-entrant jet displacing the gas occurs at a lower frequency ($St_{H,rej} \lesssim 0.1$). 
$St$ based on the time taken by the re-entrant jet to impinge on the wedge base (using $1/t = \overline{u}_{rej}/L_{c}$) matches the shedding frequency estimated from the FFT of re-entrant flow-induced gas ejection very well as shown by grey filled-star markers in figure \ref{fig:rej-spacetime}(\textit{d}). This suggests a correlation between the re-entrant jet and gas ejection. Additionally, it is observed that $St_{H,rej}$ decreases linearly: $St_{H,rej} = 0.2-2.375C_{qs, in}$ as $C_{qs}$ increases as shown by a grey dashed line in figure~\ref{fig:rej-spacetime}(\textit{d}). This empirical relation is invoked in Appendix \ref{AppC} for constructing a simple model for gas ejection.  

\begin{figure}
    \centering
    \includegraphics[width=0.94\textwidth]{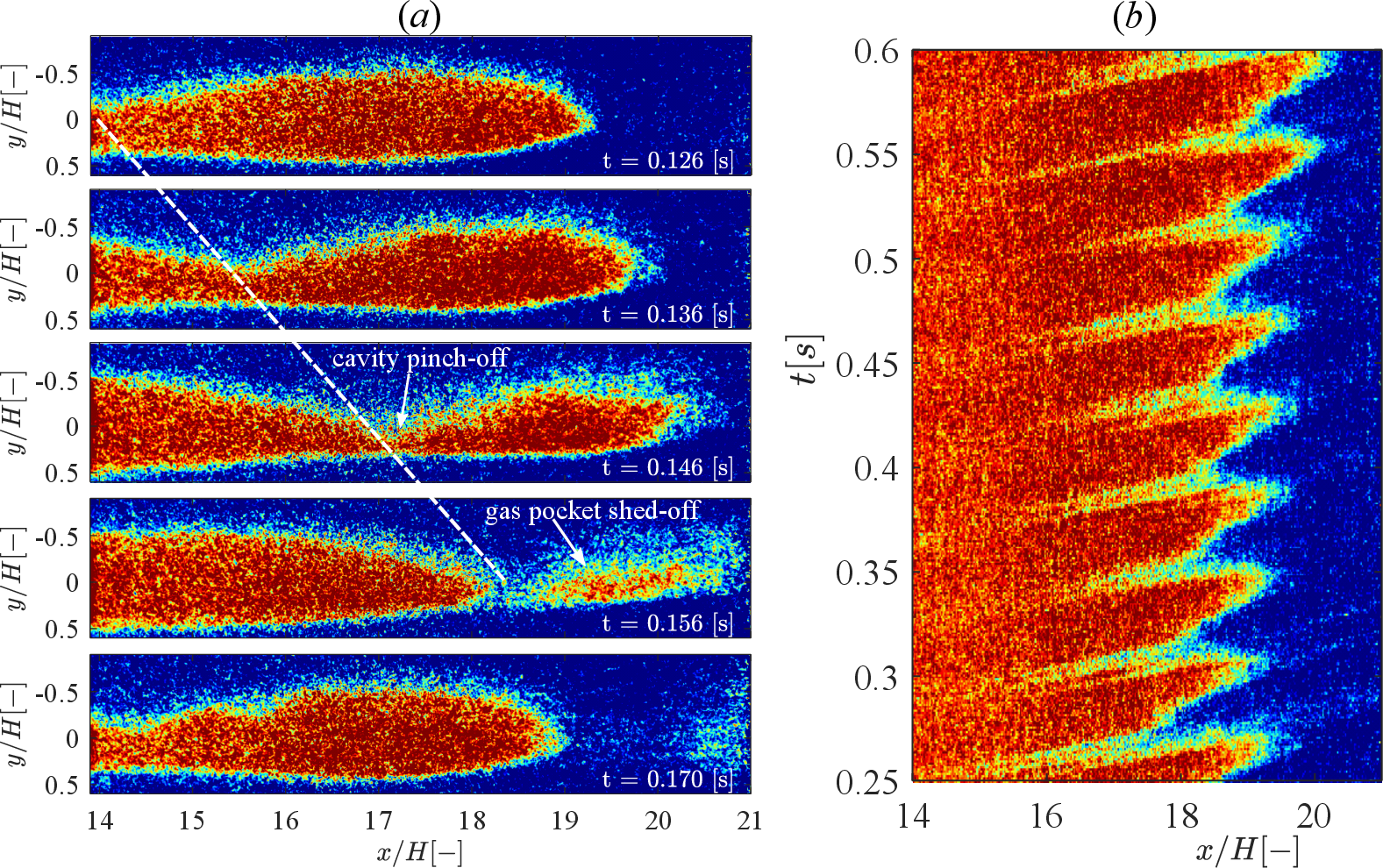}
    \caption{(\textit{a}) X-ray flow visualisation of cavity closure corresponding to the LC regime at $Fr = 5.79$, $C_{qs} = 0.069$. X-ray images were taken through the thicker and denser PVC walls of the test section leading to greatly reduced SNR. As such, the void fraction cannot be accurately estimated. However, the shape of the cavity can be visualised. The white dashed line shows the travelling wavefront leading to the gas pocket shed-off at the closure. \textcolor{black}{(\textit{b}) $x-t$ diagram at $y \approx 0$ for long cavity closure showing multiple gas shedding cycles.}}
    \label{lc_closure} %see 10/01/2023 F2
\end{figure}

\textit{Long cavities (LC)}: 
\label{closure}
Closure of LC cavities was not observed using high-speed images as it spanned beyond the FOV. Nevertheless, X-ray-based flow visualisation was performed through the thicker and denser PVC walls of the test facility at FOV-2, as shown in figure~\ref{fig:setup}. This led to a low signal-to-noise ratio in void fraction allowing only qualitative cavity visualisation. From these qualitative visualisations, the following observations were made: long cavities do not exhibit a re-entrant flow-like closure. The gas appears to be ejected out of the cavity closure via inertial wave-type instability, as shown in figure \ref{lc_closure}(\textit{a}). 
\textcolor{black}{The gas pocket shed off at the closure is shown by $x-t$ in figure  \ref{lc_closure}(\textit{b}).}
The wave is seen to be convecting on the cavity interface (see supplementary movie S9) with a velocity close to inflow velocity ($U_{0}$). 
This leads to cavity pinch-off and a gas pocket shed-off with a non-dimensional frequency ($St_{H}$) of about  0.19. This is in close agreement with \cite{Shao2022}, who reported an average $St$ of 0.20 for oscillatory cavities in an axisymmetric cavitator. With an even further increase in $C_{qs}$, the cavity lengths do not increase substantially, rather, the amplitude of oscillation increases. The size of the gas pocket shed-off is expected to increase to allow higher gas ejection. The similarity of the observed closure of the long cavity with a twin-branched cavity is an open question and will be investigated in a future investigation.

\subsection{Compressibility effects}
In natural cavitating flows, the compressibility of the medium plays a significant role in dictating cavity dynamics \citep{Ganesh2016, Wu2021}. \textcolor{black}{Moderately high} vapour fractions ($\sim$ 0.4-0.7) in combination with low pressure in the vapour cavity (of the order of vapour pressure) reduces the speed of sound in this bubbly mixture to $\mathcal{O}(1)$ ms$^{-1}$. This can result in high cavity Mach numbers, giving rise to the propagating bubbly/condensation shock fronts as observed experimentally by \cite{Ganesh2016, Gawandalkar2024} and numerically by \cite{Budich2018}. 
In ventilated cavities, the void fraction can attain values close to 0.5, which coincides with the minimum speed of sound. This can lead to a high Mach number, especially in REJ cavities due to the entrainment of liquid inside the cavity (resulting in low $\alpha$)
and the presence of re-entrant jet fronts (see figure~\ref{fig:rej-timeseries-xray}). 
Hence, the effects of compressibility need to be examined for such cavities.
The direct void fraction ($\alpha$) and pressure measurements ($P_{c}$) allow us to estimate the speed of sound in the binary mixture within the cavity ($c_{m}$), using equation \ref{eq:Speed_of_sound} \citep{Brennen1995}, where $k$ is the polytropic constant of the mixture. In equation \ref{eq:Speed_of_sound} it is assumed that the cavity is composed of liquid (subscript $l$) and gas (subscript $g$) with negligible surface tension effects, bubble dynamics, and mass transfer between phases. The Mach number is then estimated with equation \ref{eq:mach}, where the velocity scale $u$ is either the inflow velocity ($U_{0}$) or the velocity of the re-entrant jet front ($\overline{u_{rej}}$). 
\begin{figure*}
    \centering
    \includegraphics[width=1.0\textwidth]{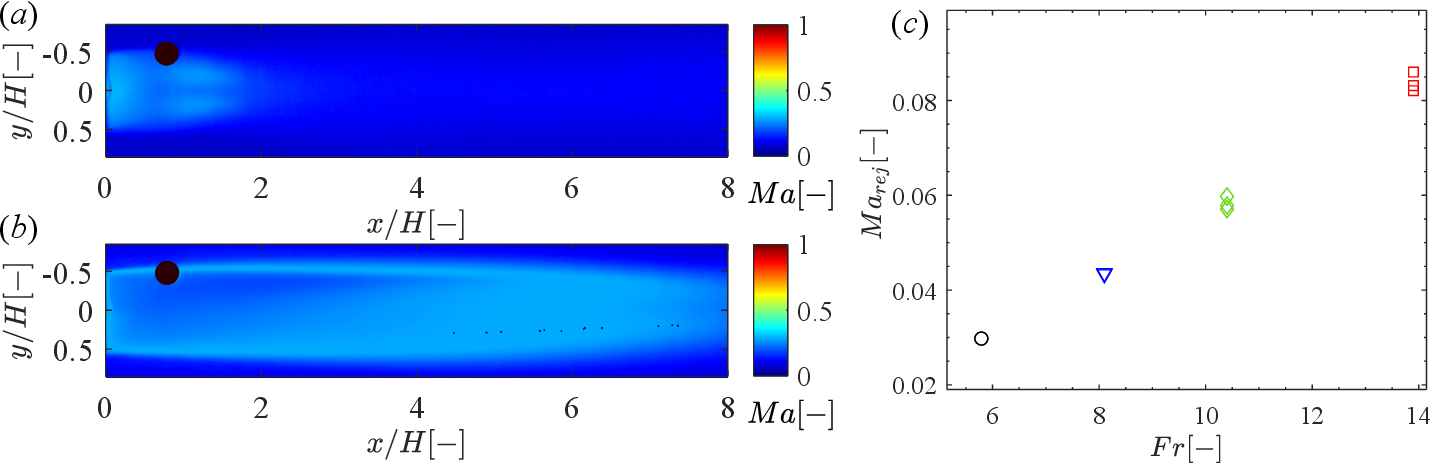}
    \caption{Mach number ($Ma$) based on averaged void fraction fields ($\overline{\alpha}$) and incoming velocity ($U_{0}$) at (\textit{a}) $Fr$=13.89 $C_{qs}$= 0.029, (\textit{b}) $Fr$=13.89 $C_{qs}$= 0.0617, (\textit{c}) Mach number of the re-entrant jet front at different $Fr$ and $C_{qs}$. }
    \label{fig:mach-field}
\end{figure*}
\begin{equation}
    \label{eq:Speed_of_sound} 
 c_{m} =  \Bigl\{ \frac{\alpha_{}}{kP_{c}} [\alpha_{} \rho_{g} + (1-\alpha_{}) \rho_{l} ]\Bigl\} ^ {-\frac{1}{2}},   
\end{equation}

\begin{equation} \label{eq:mach}
    Ma = \frac{u}{c_{m}}
\end{equation}
TBC and LC are filled with air ($\alpha \sim 1$) as seen by the void fraction distribution shown in figures~\ref{avg_vc}(\textit{b,d}). 
Hence, the speed of sound is expected to be high in these cavities, resulting in low Mach numbers at $u$. 
These cavities are omitted from the analyses. Further, using $U_{0}$ as a velocity scale to estimate the $Ma$ provides us with the maximum achievable values of $Ma$ for the cavity. The time-averaged void fraction fields are used to estimate the local $Ma$ for foamy and REJ cavities in figure~\ref{fig:mach-field}(\textit{a, b}). It is seen that the Mach number is consistently lower than 0.3. The estimated Mach number of re-entrant flow front ($Ma_{rej}$, based on $\overline{u}_{rej}$) is shown in figure~\ref{fig:mach-field}(\textit{c}). It increases with $Fr$, yet remains lower than 0.1. This suggests that the effect of compressibility is negligible for the ventilated cavities under consideration. Although the void fractions are sufficiently low, the pressure in the cavities is sufficiently high to maintain the speed of sound as high as 25 ms$^{-1}$. This is in contrast with natural cavities, where the speed of sound ($c_{m}$) drops to a few ms$^{-1}$ owing to low cavity pressures close to vapour pressures of a few kPa demonstrated by \cite{Gawandalkar2024}. Hence the bubbly shock fronts observed in natural cavitation flow are not observed in ventilated cavities under consideration.\par

% ========================= Formation of Supercavities ============================ %
\section{Formation of supercavities} \label{trans}
 The stable ventilated cavities having different closure type show distinct and fixed cavity lengths. However, during the transition from one closure type to another, the cavity length undergoes an abrupt increase evident in figure~\ref{fig:cavity-length}(\textit{a, b}). This abrupt increase in length is also accompanied by a change in cavity closure type. Cavities observed during the formation process are referred to as transitional cavities, and the corresponding closures as transitional closures. Such transitional cavity closures occur during the formation and collapse of supercavities which dictates the ventilation demands and the stability of the fully formed supercavities. Hence it is imperative to examine the formation of supercavities, with an emphasis on transitional closure region.
 
The formation of supercavities (twin-branched cavities and long cavities) was studied with an L-H ventilation strategy shown by a red $C_{qs}$ profile in figure \ref{fig:injection-profile}. Using time-synchronous measurements of void fraction fields ($\alpha(x, y, t)$) and the ventilation input ($\dot{Q}_{in}(t)$), change in cavity volume ($V_{g,c}(t)$) was estimated to quantify gas leakage ($\dot{Q}_{out}(t)$).
In a control volume (CV) around the cavity, the mass balance of gas inside the cavity is given by equation \ref{eq:cv-delRho}, where $\rho_{c}$ is the mass density, $V_{c}$ is the cavity volume and $\dot{Q}$ is the volumetric gas flux. The subscript $g$ denotes the gas phase, while $in$ and $out$ denote gas injection and ejection, respectively.
\begin{equation}
\centering
    \label{eq:cv-delRho}
     \pdv{\rho_{g,c}V_{g,c}}{t} = \rho_{g} \dot{Q}_{g,in} - \rho_{g} \dot{Q}_{g,out}
\end{equation}

\ref{eq:cv-delRho} can be rearranged to estimate the ratio of ejected to injected gas from the cavity expressed as $\Theta_{f}$, computed as per \ref{eq:cv-Qg*}. $\Theta_{f}$ is directly proportional to the gas ejection rate \emph{out} of the cavity and inversely proportional to the gas entrainment rate \emph{inside} the cavity.
\begin{equation}
    \label{eq:cv-Qg*}
      \Thetaup_{f} = \frac{\dot{Q}_{g,out}}{\dot{Q}_{g,in}} (t) = 1 - \frac{1}{\dot{Q}_{g,in}(t)} \pdv{V_{c}(t)}{t}.
\end{equation} 
In \ref{eq:cv-Qg*}, $\dot{Q}_{g,in}(t)$ is the prescribed injection profile, and $V_{c}(t)=\iint_{{CV}}\alpha(x,y,t)W\dd{x}\dd{y}$ is the estimated cavity volume from the X-ray-based void fraction fields. The control volume is a rectangle between $y/H=-0.65$ and $0.65$ for the full extent of the X-ray FOV. Note that $\Theta_{f}$ estimation is limited to snapshots where the entire cavity is within the X-ray FOV. The estimated instantaneous cavity volume ($V_{c}(t)$) is smoothed using a cubic smoothing spline with $p=0.99$ \citep{deboor78}. Quasi-stable cavities are expected to satisfy $\Theta_{f}\approx 1$, i.e. gas injection is in equilibrium with gas ejection. The formation of a twin-branched cavities (TBC) at $Fr \approx$ 2.08, 4.17 and long cavity (LC) at  $Fr \approx 10.42$ are discussed separately in \ref{unstablecavities} and \ref{lcform}, respectively. \par

\subsection{Formation of twin-branched cavity (TBC)} \label{unstablecavities} 
\begin{figure*}  
    \centering 
    \includegraphics[width=0.95\textwidth]{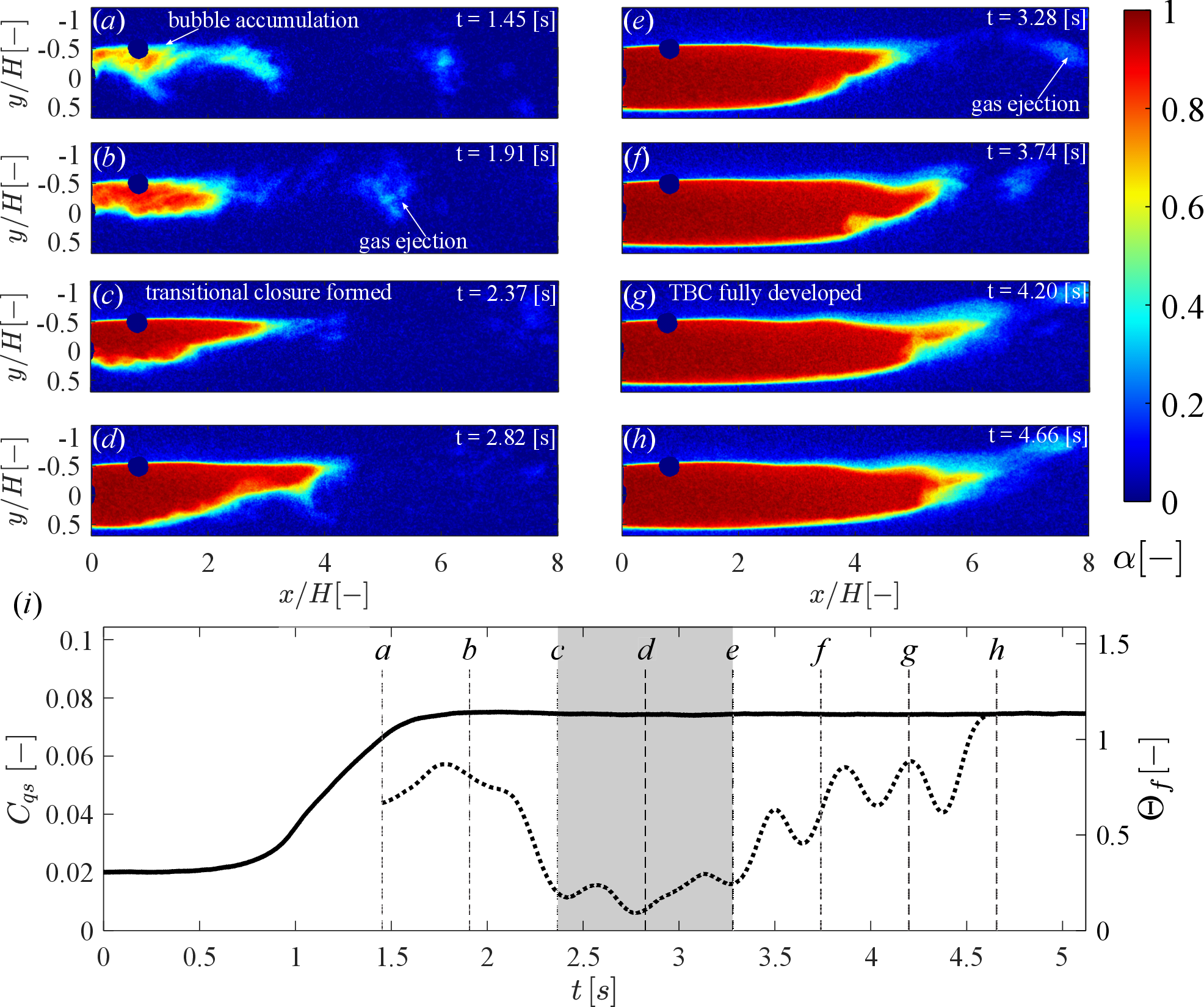}
    \caption{ (\textit{a})-(\textit{h}) X-ray densitometry snapshots of TBC formation at $Fr= 2.08$ and final $C_{qs} = 0.075$. (\textit{i}) The temporal variation of the ventilation coefficient, $C_{qs}$ (solid line), and instantaneous ejection ratio, $\Theta_{f}$ (dotted line). The grey region indicates the transitional closure region.}
    \label{fig:form:tbc-low}
\end{figure*}

Figure~\ref{fig:form:tbc-low} shows the formation of a TBC based on time-resolved X-ray measurements at $Fr = 2.08$, $C_{qs}=0.075$. 
Initially, small bubbles accumulate near the top of the wedge as shown in figure~\ref{fig:form:tbc-low}(\textit{a}--\textit{b}). 
The gas ejection ratio, $\Theta_{f}$, decreases with increasing $C_{qs}$ and the cavity begins to expand towards the bottom of the wedge, as shown in figure \ref{fig:form:tbc-low}(\textit{b}--\textit{d}). The decrease in $\Theta_{f}$ can be attributed to the formation of transitional closure leading to the increased gas entrainment to form a supercavity. As the cavity height increases to reach near-wedge height ($H$), it rapidly increases in length due to the formed closure with no observable re-entrant flow (figure~\ref{fig:form:tbc-low}(\textit{c--e})). This coincides with $\Theta_{f}$ attaining a near-constant value of 0.2 as shown in the grey region in figure~\ref{fig:form:tbc-low}(\textit{i}). It is important to note that the cavity is nominally two-dimensional during the growth or transition phase. During cavity growth, $\Theta_{f}$ exhibits slight oscillatory behaviour due to the gas ejection, via cavity pinching at the closure. Figure~\ref{fig:form:tbc-low}(\textit{e--g}) shows the formation of cavity branches along the wall, a characteristic of twin-branched cavities discussed earlier. Although oscillatory, $\Theta_{f}$ increases in (\textit{e}) to (\textit{g}) from 0.2 to 1. This suggests that the cavity is approaching a fully developed state, i.e. $\dot{Q}_{g,in}$ is in equilibrium with $ \dot{Q}_{g,out}$. The cavity at such low $Fr$ exhibits a significant degree of camber and the closure is marked by a wave--type instability responsible for gas ejection. The cavity formation dynamics are shown in the supplementary movie S10.  \par

Figure~\ref{fig:form:tbc-high} shows the formation of a TBC at $Fr = 4.2$ and $C_{qs}=0.096$. The formation dynamics are different due to the increased effect of flow inertia.
Initially, there is a dispersed, foamy wake behind the wedge, and gas ejection due to the vortex shedding is seen in figure~\ref{fig:form:tbc-high}(\textit{a--c}), similar to figure~\ref{fig:form:tbc-low}(\textit{a}--\textit{b}). With an increase in injection rate ($Q_{g,in}$), $\Theta_{f}$, although oscillatory, decreases to about 0.4 near \textit{d} in figure~\ref{fig:form:tbc-high}(\textit{i}), suggesting increased gas entrainment. The observed oscillation in $\Theta_{f}$ is attributed to gas ejection by span-wise vortex streets, as shown in figure~\ref{fig:form:tbc-high}(\textit{b}) and (\textit{c}). The above-mentioned features are visualised in the supplementary movie, S11. 
The decrease in $\Theta_{f}$ in figure \ref{fig:form:tbc-high}(\textit{d--h}) coincides with the formation of transitional closure and relatively rapid gas accumulation near the top of the wedge, increasing the cavity length. Interestingly, during the cavity growth (after the formation of transitional closure), $\Theta_{f}$ remains almost constant at about 0.4, shown by the grey region in figure~\ref{fig:form:tbc-high}(\textit{i}). This behaviour is similar to the previous case of $Fr$ = 2.08, where $\Theta_{f}$ maintained a near-constant value 0.2. 
As the cavity grows, the upper cavity interface curves downwards as evident in (\textit{d}).
This downward curvature near the cavity closure is in contrast to the upward curvature seen at the lower $Fr$ case shown in figure~\ref{fig:form:tbc-low}.
The downward curvature is accompanied by a fixed length re-entrant flow ($l_{rej,max}$) at the cavity closure as visualised in figure~\ref{fig:form:tbc-high}(\textit{e--h}) by low void fraction region ($\alpha \approx$ 0.5) in the lower half of the cavity.  
As the cavity grows in length, the re-entrant flow front recedes further away from the wedge base as shown in figure~\ref{fig:form:tbc-high}(\textit{d--h}). The transitional cavities are nominally two-dimensional in contrast to fully formed TBC with three-dimensional closure. Furthermore, the structure of the transitional closure is seen to be dependent on the flow inertia ($Fr$, $Re_{H}$), addressed in subsection  \ref{pressure_gradient_closure}.
\begin{figure}  %see 13/01/2023 F15
    \centering
    \includegraphics[width=0.95\textwidth]{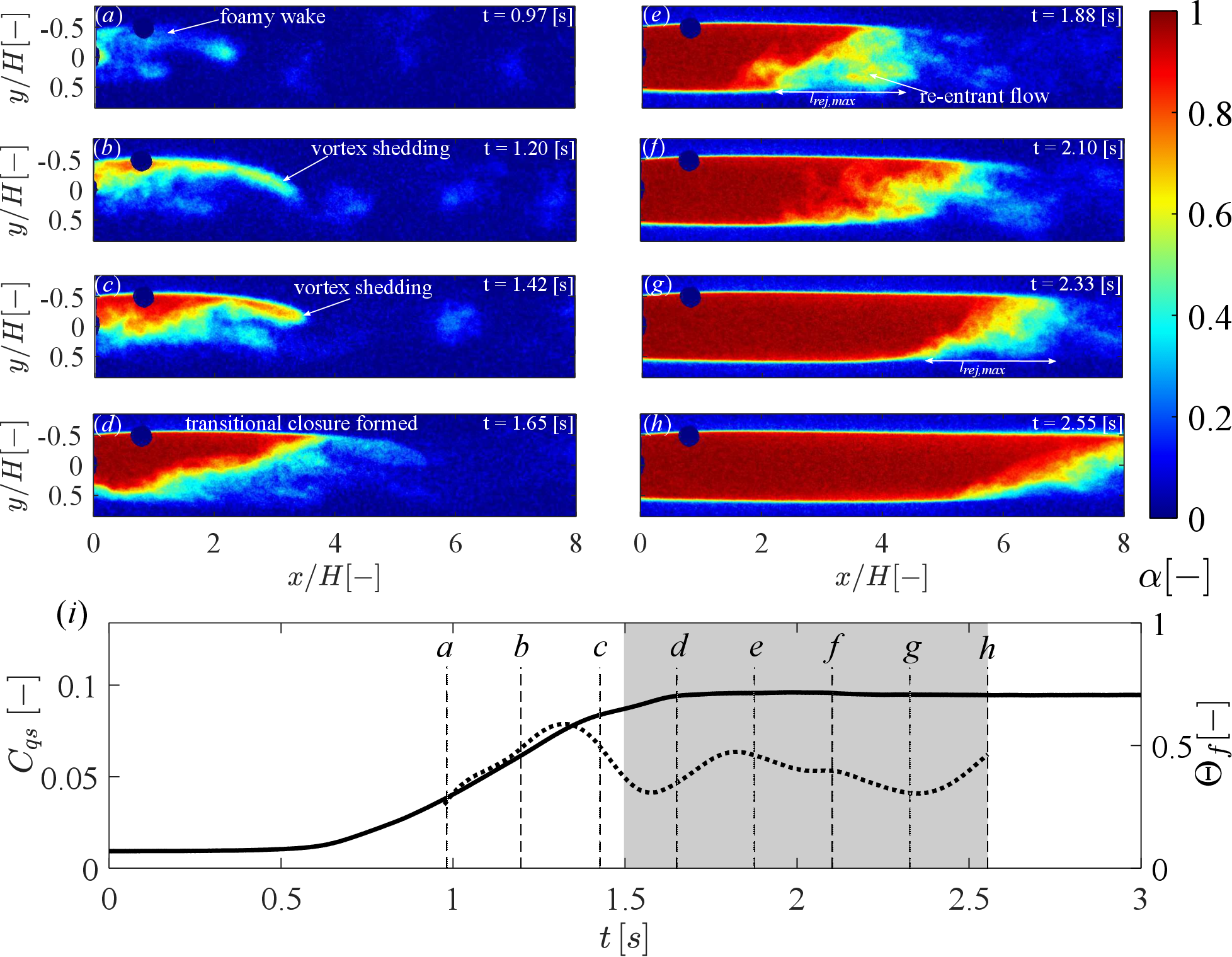}
    \caption{ (\textit{a})-(\textit{h}) X-ray densitometry snapshots of TBC formation at $Fr= 4.17$ and $C_{qs} = 0.096$. (\textit{i}) The time-dependent ventilation coefficient, $C_{qs}$ (solid line), and instantaneous ejection ratio, $\Theta_{f}$ (dotted line).}
    \label{fig:form:tbc-high}
\end{figure}
\begin{figure*}  %see 17/01/2023 F23
    \centering
    \includegraphics[width=0.95\textwidth]{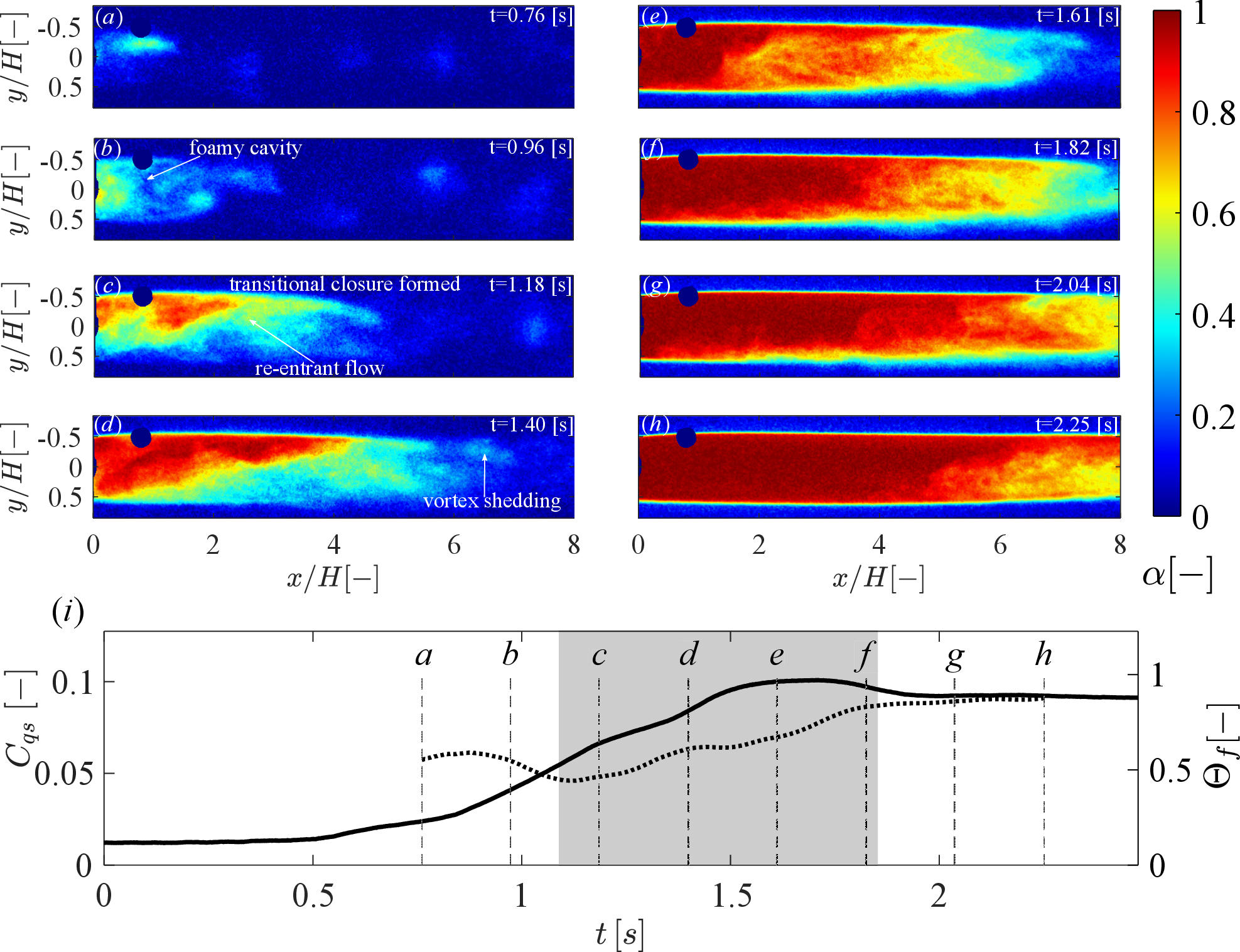}
    \caption{ (\textit{a})-(\textit{h}) X-ray densitometry snapshots of LC formation at $Fr= 10.4$ and $C_{qs} = 0.090$. (\textit{i}) temporal variation of the ventilation coefficient, $C_{qs}$ (solid line), and instantaneous ejection ratio, $\Theta_{f}$ (dotted line). The grey region indicates the transitional closure region.}
       \label{fig:form:lc}
%\end{figure*}
 %\begin{figure*} 
 \vspace{2mm}
     \centering
     \includegraphics[width=0.97\textwidth]{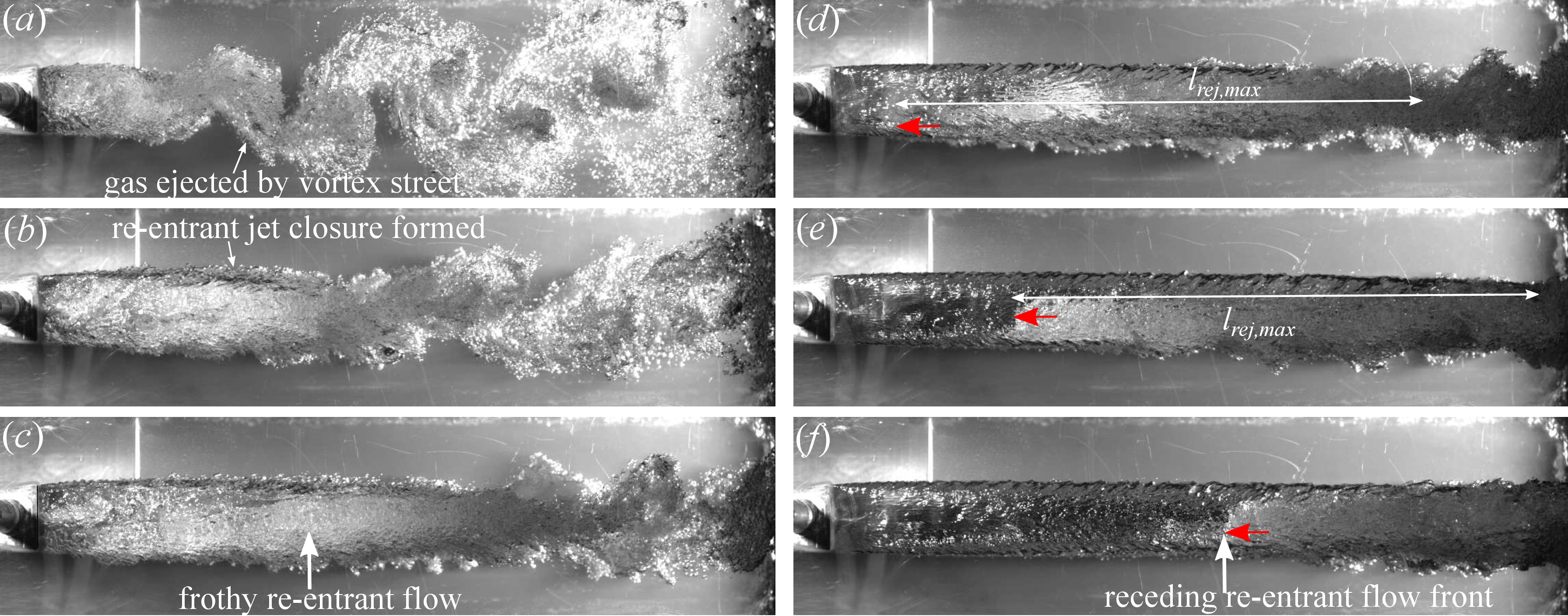}
     \caption{High-speed visualisation of LC formation (gas entrainment, liquid re-entrainment and gas ejection) at $Fr$ = 10.4, $C_{qs}$  = 0.090. The red arrow indicates the re-entrant flow front as the cavity length increases. }
     \label{fig:form:lc:hs}
 \end{figure*}

\subsection{Formation of long cavity (LC)} \label{lcform}
 Figure~\ref{fig:form:lc} shows the formation of long cavities at $Fr= 10.4$ and $C_{qs} = 0.090$. The corresponding high-speed visualisations having a larger FOV shown in figure \ref{fig:form:lc:hs} facilitate a complete observation of flow structures at the closure region.
 As gas injection begins, a foamy cavity (FC) is formed, as shown in figures \ref{fig:form:lc}(\textit{a--b}). As $C_{qs}$ is increased, transitional closure is formed and supercavity formation begins near \textit{c}. Thus foamy cavity transitions to REJ cavity upon the formation of transitional closure with a re-entrant flow at \textit{c}. $\Theta_{f}$ reaches a minimum value of 0.45, suggesting increased entrainment of injected gas in the wake leading to an increase in cavity length as shown in \ref{fig:form:lc}(\textit{i}).
  This results in a supercavity with a re-entrant flow spanning the entire cavity length (i.e. more than $4H$) and impinging on the wedge base at \textit{d}. 
 The gas is ejected via re-entering liquid flow and spanwise vortices visualised by Von K\'arm\'an vortex streets visualised in figure \ref{fig:form:lc:hs}(\textit{b,c}).
 Figure \ref{fig:form:lc}(\textit{e--h}) shows that with a further increase in $C_{qs}$, the re-entrant jet has assumed its maximum length similar to the $Fr \approx 4.17$ case. As the REJ cavity begins to transition to LC, the re-entrant flow front can no longer impinge on the wedge base. Instead, it recedes away from the wedge base as the cavity length increases as also reported by \cite{Barbaca2017}. The transition of REJ cavity to LC is shown by supplementary movie, S12, and in figure \ref{fig:form:lc:hs}(\textit{d--f}), where the receding re-entrant jet front is indicated by a red arrow. During the transition, the length of the re-entrant flow remains constant ($l_{rej, max}/H \approx$ 8.5) as its front recedes further away from the wedge base with the increase in cavity length.
Interestingly, as the cavity length increases (after the formation of closure at \textit{c}), $\Theta_{f}$ does not maintain a constant value, rather it increases from 0.45 to 0.85 (see the grey region in figure~\ref{fig:form:lc}(\textit{i})), suggesting larger gas ejection rates in comparison to previous cases of low $Fr$. \par
The different trends in $\Theta_{f}$ with $Fr$ can be explained by the distinct closure types transitional cavities attain during the formation of TBC and LC: During the formation of TBC, the transitional cavity has a wave-type instability or a mild re-entrant flow with vortex shedding, resulting in a smaller gas ejection rate. 
However, during the formation of LC, the closure has a prominent re-entrant flow entering the cavity, resulting in a larger gas ejection.
A prominent re-entrant flow at the closure seems to contribute to a significantly larger gas ejection than wave-type instability or vortex shedding.
This larger gas ejection rate likely leads to reduced gas entrainment, and hence stunted length of REJ cavities in figure \ref{fig:cavity-length}(\textit{c}).
This is in line with the DIH measurements of \cite{Shao2022}, who reported a large instantaneous gas ejection rate for cavities with re-entrant jet closure. 
\textcolor{black}{This is further corroborated by the higher collapse rate of established supercavities at higher $Fr$ cases having a prominent re-entrant flow at the closure (not shown here).}
The dynamics of re-entrant flow at the cavity closure with increasing $Fr$ is discussed in the next subsection. \par

\subsection{Effect of flow inertia on re-entrant jet cavity closure} \label{pressure_gradient_closure}
\begin{figure}
    \centering
    \includegraphics[width=0.95\textwidth]{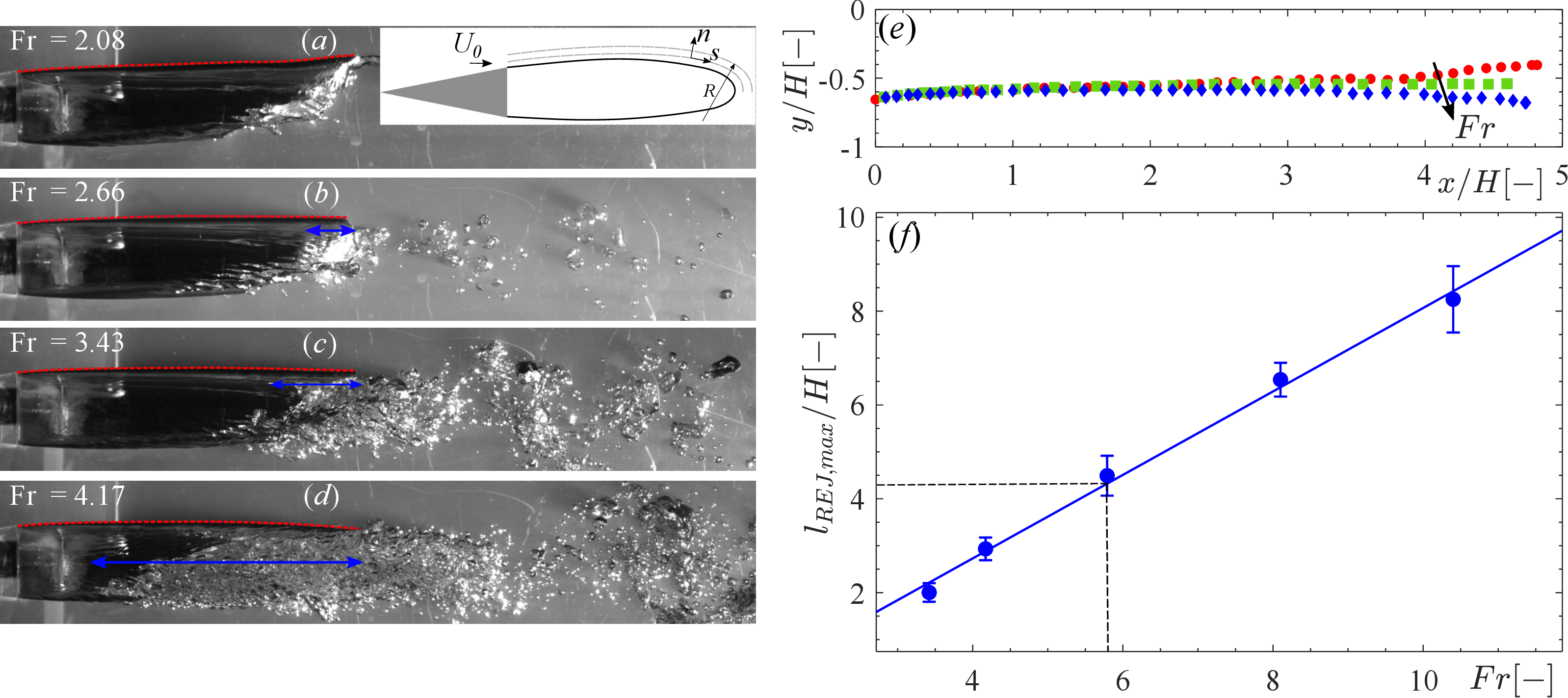}
    \caption{ (\textit{a})-(\textit{d}) Transitional cavity closure at a fixed cavity length seen at low $Fr$ range during the transition from FC to TBC. The blue marking shows the length of the re-entrant flow, while the red dashed lines show the upper cavity interface. (\textit{e}) The upper cavity interface as a function of $Fr$ shows that the cavity interface curves downwards with an increase in $Fr$. (\textit{f}) The maximum re-entrant flow length ($l_{rej, max}/H$) as a function of $Fr$. The solid blue line is a linear fit: 0.85$Fr$--0.73. The black dashed line shows the minimum length of the stable REJ cavity observed in this study at $Fr \approx$ 5.7.}
    \label{unstable_cav}
\end{figure}

During the supercavity formation process, the length of the liquid re-entrant flow attains a maximum, constant value, denoted $l_{rej, max}$ as also observed by \cite{Barbaca2017}. After $l_{rej, max}$ is attained, the re-entrant flow front cannot reach the wedge base, i.e. $l_{rej, max}$ is shorter than the cavity length. The existence of constant length re-entrant flow at the cavity closure during the supercavity formation is observed for $Fr \geq$ 3.43. 
$l_{rej, max}$, indicated by blue arrows in figure \ref{unstable_cav}(\textit{a--d}) can be used as a proxy for the intensity of the re-entrant flow driven by a pressure gradient. $l_{rej, max}$ is plotted for a range of $Fr$ for a fixed cavity length in figure \ref{unstable_cav}(\textit{f}). It is seen to scale linearly with $Fr$: $l_{rej, max}/H \sim$ 0.85$Fr$ -- 0.73, suggesting that the re-entrant flow is getting stronger with flow inertia (see figure \ref{unstable_cav}(\textit{f})). Furthermore, $l_{rej, max}$ is seen to be independent of the ventilation coefficient, $C_{qs}$ for a given $Fr$. This is in consistent with \cite{Barbaca2017, Barbaca2019} who showed a similar behaviour of re-entrant jet length in natural and ventilated cavity flows behind a fence. It is worth noting that the increase in $Fr$ is accompanied by a stronger curvature of the upper cavity interface, resulting in a smaller radius of curvature ($R$) (see figure \ref{unstable_cav}(\textit{e}) and inset in \textit{a}). For instance, $Fr=2.08$ shows a slight upward curvature, while as $Fr$ increases, stronger downward curvature is exhibited by the upper cavity interface as shown in figure \ref{unstable_cav}(\textit{e}). 
This results in a smaller radius of curvature leading to a stronger adverse pressure gradient at the cavity closure as demonstrated by the 1-D Euler's equation in streamline coordinate systems (\ref{dp}). Here, $n$ is a direction normal to the radius of curvature in a streamline coordinate system, and $R$ is the radius of curvature of the cavity closure (see inset in figure~\ref{unstable_cav}\textit{a}). A strong pressure gradient at the closure drives upstream, re-entrant flow from the closure region as also observed in natural cavitating flows \citep{Callenaere2001, Gawandalkar2022}. 
 \begin{equation}
    \frac{\partial p }{\partial n} \sim \rho \frac{U_{0}^{2}}{R} 
    \label{dp}
 \end{equation}
%Interestingly, such re-entrant jets are not reported in the study of \cite{Wu2021} dedicated to natural cavities in the same wedge geometry. 
The length or the intensity of the re-entrant jet is suspected to play a significant role in determining the geometry (length) of the stable  REJ cavities: Stable REJ cavities are seen to span over 4--8 wedge height(s) ($H$). At $Fr$ = 5.79, $l_{rej, max}$ is 4.2$H$, which is the minimum length of the stable REJ cavity (see figure~\ref{fig:cavity-length}(\textit{b})). This then coincides with the critical $Fr$ at which stable REJ cavities start to exist, shown by the black dashed line in figure \ref{unstable_cav}(\textit{f}). This suggests an effect of wake-flow interaction with the gas entrainment on the formation of re-entrant jet cavities. Furthermore, a re-entrant flow at the closure is likely to reduce the in-plane ($x-y$) cavity oscillations observed in TBCs and LCs. It is recognised that above the critical Froude number of 5.7, the Reynolds number ($Re_{H}$) of the flow is a more suitable parameter. However, the effect of $Re_{H}$ and $Fr$ are coupled in this study and will be segregated in future experiments with different wedge geometries. Moreover, $l_{rej, max}$ possibly dictates the gas ejection out of the cavity when REJ cavities transition to long cavities. Consequently, if the ventilation input ($C_{qs}$) can overcome the gas ejection due to the re-entering liquid flow, the REJ cavity transitions to a long cavity. This is verified quantitatively using a simple gas ejection model based on empirical relations obtained in this study as detailed in Appendix \ref{AppC}.
% ============================ Ventilation Hysteresis ============================ %
\section{Ventilation hysteresis}
In the previous section(s), ventilated cavities were established using an L-H ventilation strategy which involved increasing the ventilation coefficient ($C_{qs}$) from zero to a given $C_{qs}$. In this section, the ventilated cavities formed by reducing $C_{qs}$ \textit{after the formation of supercavities} are discussed. This ventilation strategy is referred to as H-L as shown by the black $C_{qs}$ profile in figure~\ref{fig:injection-profile}. The details of the experimental procedure and ventilation strategy are elaborated in subsection \ref{procedure}.
\begin{figure}
    \centering
    \includegraphics[width=0.98\textwidth]{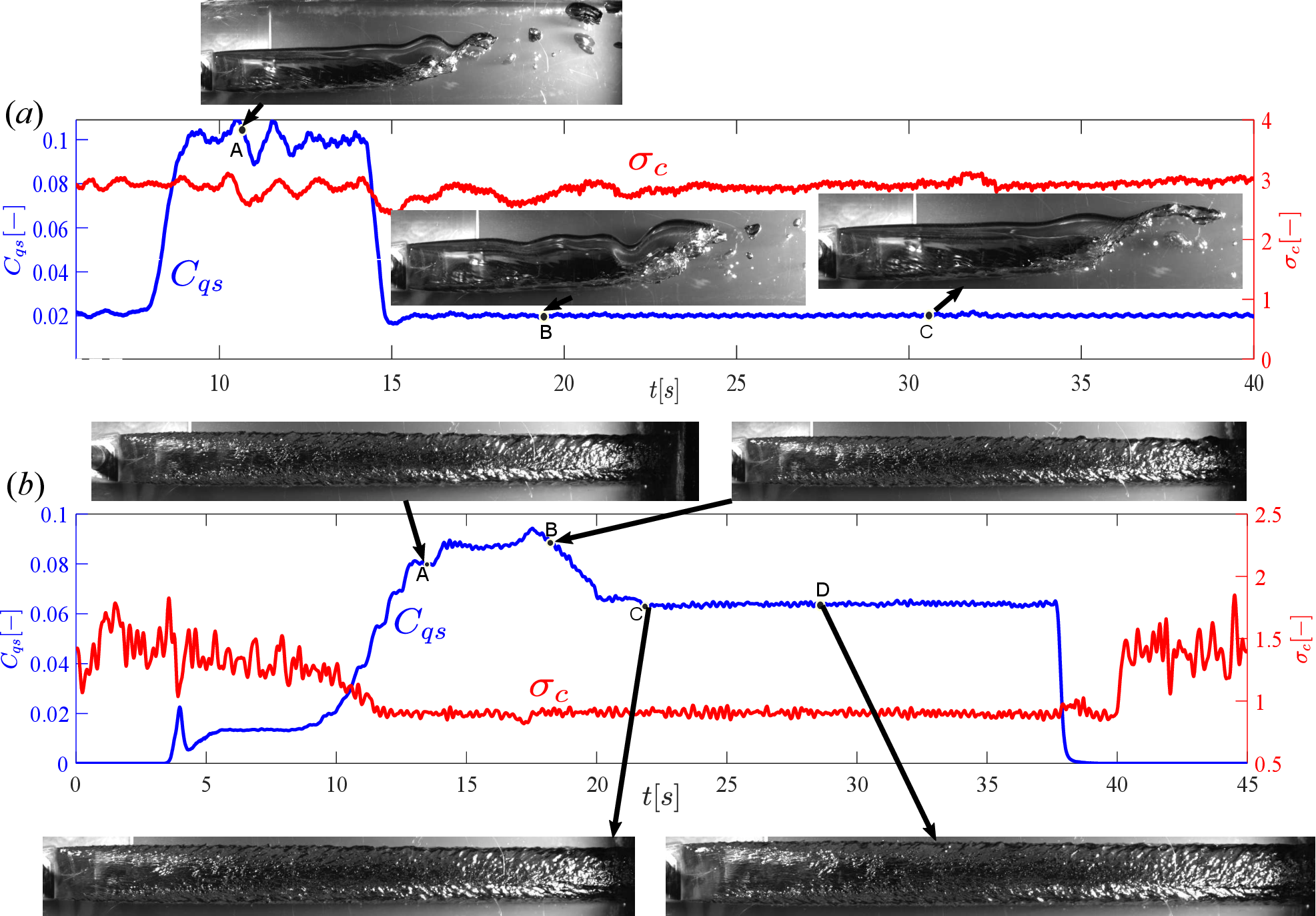}
    \caption{Time signals of $C_{qs}$ and $\sigma_{c}$ showing ventilation hysteresis using H-L ventilation strategy for (\textit{a}) $Fr$ = 2.08: TBC is formed at A and can be maintained upon $C_{qs}$ reduction at B and C. (\textit{b}) $Fr$ = 10.41: LC is formed at A and B, and can still be maintained upon $C_{qs}$ reduction at C and D. The corresponding high--speed optical visualisations of the supercavity are shown.} 
    \label{fig:hyst_comapare}
\end{figure}
\subsection{Supercavities formation with H-L ventilation strategy}
Figure \ref{fig:hyst_comapare}(\textit{a}) and (\textit{b}) shows a representative case of supercavities formed with H-L ventilation strategy for TBC and LC, respectively. The temporal variation of $C_{qs}$ is shown in blue, and the cavity pressure variation expressed as $\sigma_{c}$ is shown in red. In all cases considered, it is ensured that the cavity closure is completely formed, i.e., TBC at $Fr \approx$ 2.08 and LC at $Fr \approx$ 10.4 before reducing the ventilation coefficient to a prescribed $C_{qs}$ value. \par

At $Fr \approx 2.08$, the $C_{qs}$ is initially set to 0.1 to ensure that the twin-branched cavity is completely formed, as shown by `A' in figure~\ref{fig:hyst_comapare}(\textit{a}). Points `B' and `C' represent the region of reduced $C_{qs} \approx$ 0.019. Interestingly, it is observed that the twin-branched cavity persists at a much lower $C_{qs}$ with the H-L strategy compared to the L-H strategy . At such a flow condition ($Fr$ = 2.08, $C_{qs}$ = 0.019), foamy cavities were observed in the L-H ventilation strategy as shown in figure~\ref{fig:regime}.\par
At a higher $Fr$ ($\approx$ 10.4), $C_{qs}$ is initially increased to 0.088 such that a long cavity is formed as shown by `A' and `B' in figure \ref{fig:hyst_comapare}(\textit{b}). The reduction of $C_{qs}$ to 0.062 resulted in the persistence of long cavity as shown by `C', `D' in figure~\ref{fig:hyst_comapare}(\textit{b}). At the same flow conditions, re-entrant jet cavities were observed using the L-H strategy in the regime map of figure~\ref{fig:regime}. It is observed that upon formation of a supercavity (TBC and LC), the cavitation number ($\sigma_{c}$) remains relatively unchanged, with the reduction of the ventilation coefficient ($C_{qs}$), as shown by the red line in figure~\ref{fig:hyst_comapare}.
This is consistent with the previous observations of \cite{Karn2016} and \cite{arn} in other cavitator geometries. It is worthwhile to note that the previous studies of \cite{Karn2016} observed a change in cavity closure with the reduction in $C_{qs}$, with no effect on cavity length ($L_{c}$) and cavitation number ($\sigma_{c}$). In this study, a change in the closure is \emph{not} observed for TBCs, however, the closure of long cavities is out of the optical FOV and needs further investigation. Nevertheless, qualitative observation with X-ray imaging in FOV-2 hints that closure remains unchanged. Furthermore, we observe that in-plane cavity oscillations are `damped-down' upon $C_{qs}$ reduction. 
 
\begin{figure}
    \centering
    \includegraphics[width=1.0\textwidth]{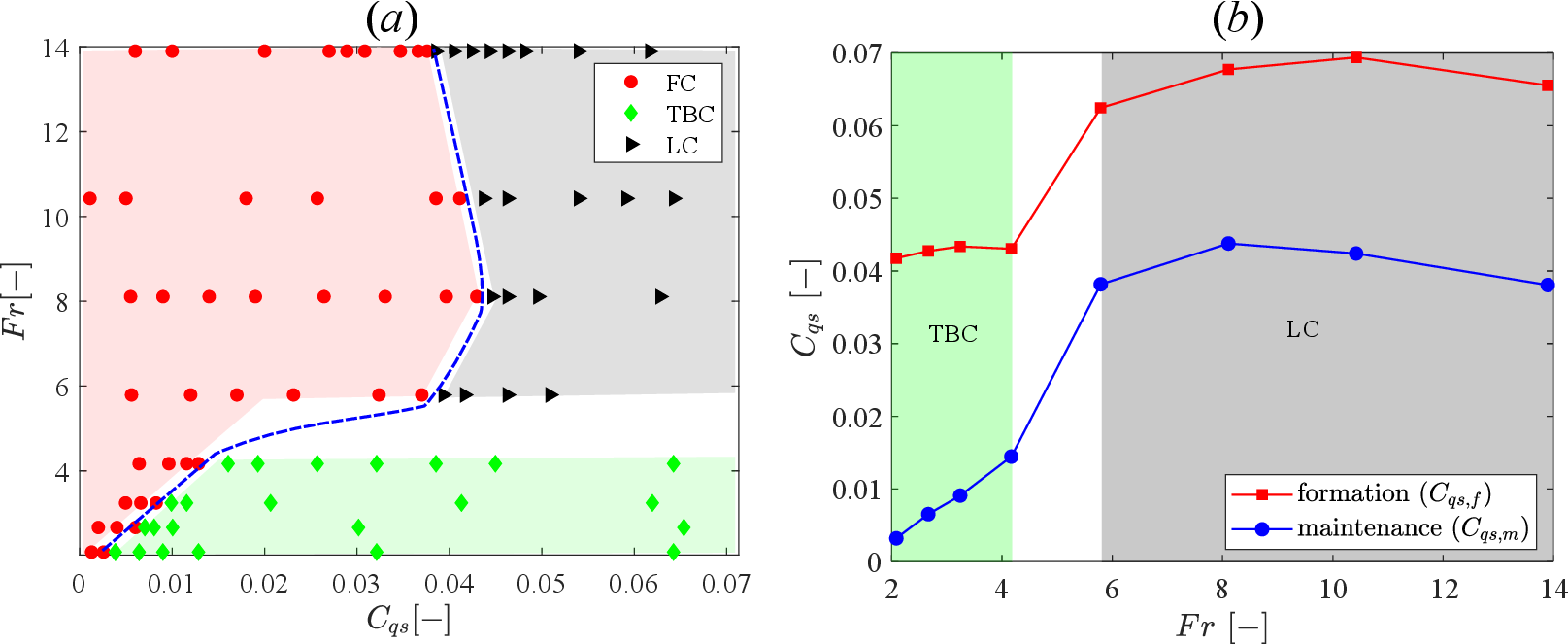}
    \caption{(\textit{a}) Cavity regime map resulting from decreasing ventilation from a fully developed supercavity initial condition (H-L ventilation strategy). The blue dashed line shows the (approximate) $C_{qs}$ limit where the supercavities are no longer maintained. (\textit{b}) Ventilation coefficient required to form and maintain cavities at different $Fr$. The formation (red) line is the minimum $C_{qs}$ required to establish a given cavity closure (TBC for $Fr \lesssim  4.2$, LC for $Fr \gtrsim 5.7$). The maintenance line (blue) is the minimum $C_{qs}$ required to maintain a given closure, below which the cavity rapidly transitions to a foamy cavity.}
    \label{fig:hysteresis}
\end{figure}

\subsection{Regime map with H-L ventilation strategy}
 A systematic variation of $Fr$ and $C_{qs}$ using H-L ventilation strategy resulted in the second regime map shown in figure~\ref{fig:hysteresis}(\textit{a}). This regime map significantly differs from that based on the L-H ventilation strategy shown in figure \ref{fig:regime}.
 At low $Fr$, it is seen that the transition boundary from FC to TBC, denoted by the interface between the green and red regions in figure~\ref{fig:hysteresis}(\textit{a}), increases monotonically with $Fr$ as opposed to a constant $C_{qs}$ = 0.043 in the regime map based on the L-H strategy.
 At higher $Fr$, the transition boundary between FC to LC, denoted as the interface between the black and red regions in figure~\ref{fig:hysteresis}(\textit{a}), is similar to the FC to REJC transition in the L-H regime map (blue dotted line in figure~\ref{fig:regime}). 
 With the H-L ventilation strategy, long cavities persist at comparable $C_{qs}$ where REJ cavities were observed for the L-H ventilation strategy. Further they \textit{transition directly to FC upon $C_{qs}$ reduction}. 
Thus, for a given $Fr$ and $C_{qs}$, the cavity closure, pressure, and geometry depend on the ventilation strategy, in fact, ventilation history.
Most importantly, from the regime maps, it is clear that the ventilation required to \em{maintain} a supercavity ($C_{qs,m}$) is significantly different than the ventilation required to \em{form} a cavity ($C_{qs,f}$) as demonstrated in figure~\ref{fig:hysteresis}(\textit{b}). \par

\subsection{Formation and maintenance gas fluxes}
 The minimum amount of gas needed to maintain a supercavity of a given length at a given $Fr$ using the H-L injection strategy can now be estimated as shown in figure \ref{fig:hysteresis}(\textit{b}). Red lines represent the gas flux needed to form a supercavity of a given length, and blue lines represent the gas flux needed to maintain the established supercavity of a given length. It is evident from figure \ref{fig:hysteresis}(\textit{b}) that more gas is needed to form a supercavity when compared to maintenance for all the $Fr$ under consideration. A similar observation has been made in ventilated cavities generated behind a two-dimensional backward-facing step \citep{arn, Makiharju2013a}.  \par
The physical mechanism responsible for the observed difference in ventilation demands can be explained with the help of cavity formation dynamics, especially the cavity closure type, and the resulting gas ejection rates during the formation. In both L-H and H-L strategies, foamy cavities (FC) are identified at low $C_{qs}$, characterised by dispersed bubbles in the near-wake region with no observable closure.
Thus the role of the near-wake flow and its interaction with the gas injection process is dominant at low ventilation ($C_{qs}$) for both strategies.  
At higher $C_{qs}$, LCs or TBCs are observed for both strategies, depending on the Froude number. Long cavities and twin-branched cavities, both, have wave--type flow characteristics. 
This is in contrast to the near-wake flow dominance at low $C_{qs}$ and hence, other mechanisms become important at high injection rates, suggesting the decreased to no influence of the wake flow at higher cavity lengths. \par

The existence of REJ cavities \emph{only} during L-H strategy is again due to the stronger influence of the wake flow during the formation process, but not during the maintenance of supercavities.
During formation, the separated shear layers have a significant influence on the injected gas and this results in the formation of closure kinematics representative of REJ cavities. 
As a result, the wake-flow influences the ventilation demands for the L-H strategy due to increased gas ejection caused by the liquid re-entrant flow.
Therefore, higher ventilation ($C_{qs}$) is required to overcome this increased gas ejection to form a long cavity leading to increased $C_{qs, f}$.\par

\textcolor{black}{During maintenance, established supercavity (LC and TBC) is used as an initial condition before reducing $C_{qs}$ to a prescribed value.} The closure observed for LC and TBC does not exhibit wake characteristics. 
Instead, a wave-type closure is identified, which allows relatively less gas ejection.
To this end, the maximum supercavity length, and hence the cavity volume is set by the Froude number. 
A \emph{minimum} volume of gas is required to feed the cavity volume to maintain it.
Consequently, for a given $C_{qs}$, the cavity receives more gas flux than it requires to sustain/maintain itself. This allows $C_{qs}$ to be decreased to $C_{qs, m}$ while maintaining long cavities, i.e. TBC and LC. 
Due to a lack of quantitative X-ray-based measurements at the closure of long cavities, the exact dynamics at its closure could not be examined. Quantification of cavity lengths across the entire range $Fr$ is expected to provide more insights into this behaviour. This is being investigated in a follow-up work.

% =========================== Summary and conclusion ============================ %
\section{Summary and Conclusions} \label{summary}
Ventilated cavities in the wake of a two-dimensional bluff body are studied with time-resolved X-ray densitometry over a wide range of flow velocities, i.e. Froude Number ($Fr \approx 2-13.9$) and Reynolds Number ($Re_{H} \approx 1.6 - 11 \times10^{4}$), and gas injection rate ($C_{qs} \approx$ 0.03-0.13). The ventilated cavities are created by systematically varying $Fr$ and $C_{qs}$. This led to a regime map ($Fr$ vs $C_{qs}$) where four types of stable, fixed-length cavities are identified, each having a unique cavity closure type. \par
Foamy cavity (FC), resembling an \text{open} cavity without a well-defined closure, was observed at low $C_{qs}$, irrespective of the $Fr$ considered. In this regime, the gas entrained in the near-wake of the wedge is ejected out periodically by the  Von K\'arm\'an vortex street. Twin-branched cavities (TBC), re-entrant jet cavities (REJC), and long cavities (LC), on the other hand, have developed closures. A TBC is formed at low $Fr$ and higher $C_{qs}$ with a three-dimensional closure consisting of two branches along the walls. The gas is ejected primarily by a wave-like travelling instability that pinches the cavity at the branches.
\par
REJ cavities are observed at an intermediate $C_{qs}$ when $Fr \geq $ 5.8. REJ cavities have a strong liquid re-entrant flow entering the cavity due to a higher pressure gradient at the closure.
The re-entrant flow in these cavities is seen to span the entire cavity length.
The gas ejection is brought about by periodic spanwise vortex shedding and a strong periodic re-entrant liquid flow displacing the entrained gas. The re-entrant flow front velocity in the laboratory frame of reference is measured to be $\sim 0.3U_{0}$, for all $Fr$ and $C_{qs}$. For a given $Fr$, the length of REJ cavities increases monotonically with increasing $C_{qs}$.
With a further increase in ventilation, for $Fr \geq $ 5.8, long cavities are observed. LCs are characterised by a wave-type instability at the closure leading to periodic gas ejection. LC closure could not be measured quantitatively due to the limitation imposed by the experimental setup. Nevertheless, long cavities are suspected to have a two-dimensional closure from our qualitative flow visualisations, as also reported by \cite{Michel1984}. Long cavities deserve a thorough quantitative investigation and are considered for future investigation. 

In addition to the topological features reported, several observations from the study are summarised below. The cavity closure and the resulting gas ejection rate are seen to influence the cavity length and the pressure within the cavity. Furthermore, an observed change in cavity closure is accompanied by an abrupt increase in cavity length. It is also observed that the cavity length is not unique to a flow condition given by $Fr$ and $C_{qs}$; a cavity of a given $L_{c}$ can have \emph{different} closures. Cavity pressure, expressed as cavitation number ($\sigma_{c}$), determines the observed closure for a given cavity length. Additionally, despite low void fractions ($\sim$ 0.5) in the cavity, no significant effect of water-air mixture compressibility was observed on the cavity dynamics. This is primarily due to high cavity pressures. The supercavity lengths, when plotted as a function of  ($\sigma_{c}$), appear to follow a power law relationship except for REJ cavities. This is likely linked to the re-entering jet displacing larger amounts of gas, resulting in a larger gas ejection or low gas entrainment rate to form the cavity. Thus stable REJ cavities are \emph{stunted} in length. TBCs and LCs exhibit a maximum cavity length for a given $Fr$, implying limited permissible gas entrainment in the wake. Qualitatively, we observed that stable cavities after assuming the maximum length, start to oscillate more in the $x-y$ plane with an increase in $C_{qs}$ to eject the \emph{excess} gas.

Based on simple gas balance analysis, it is observed that upon the formation of transitional closure, the gas ejection rates decrease rather abruptly, leading to an abrupt increase in cavity length. The supercavity formation process differs at low and high $Fr$. Specifically, the normalised gas ejection during the formation at low $Fr$ is seen to be lesser than the gas ejection at high $Fr$. This is due to the different cavity closures attained by the cavity and the resulting wake kinematics. At low $Fr$, the gas is ejected out via span-wise vortices or wave instability. However, at higher $Fr$, gas is ejected out via spanwise vortices \textit{and} primarily via a re-entrant liquid flow. A prominent re-entrant flow at the closure results in higher gas leakage rates than vortex or wave-type closure.
A fixed length, re-entrant flow is seen at the closure of transitional cavities as the supercavities are being formed at $Fr \geq 3.43$. 
The length of this re-entrant flow increases with increasing flow inertia, which in turn dictates the gas ejection and the observed transition from REJ cavities to long cavities. 

It is observed that cavity closure exhibits ventilation hysteresis, i.e. for a given $Fr$, cavity closure depends on the ventilation history.
The transitional cavity closure and the resulting gas leakage during the cavity formation provide us insights into observed ventilation hysteresis. 
The ventilated cavities at low $C_{qs}$ are seen to be dominated by the interaction of wake-flow with gas injection.
However, at higher $C_{qs}$, longer cavities (TBCs and LCs) exhibit no influence of wake interaction, instead, they have wave-type characteristics.
For a given $C_{qs}$, a stronger wake interaction results in a higher gas ejection rate in REJ cavities in comparison to wave-type closure in TBC and LC.
Due to the different gas ejection characteristics at different closures, ventilation demands are dependent on the closure history.
This explains why supercavities can be maintained/sustained at almost half the gas flux required to form/establish the supercavity.

The findings of the present study suggest that the ventilation strategy is paramount for the efficient formation and maintenance of ventilated supercavities. This understanding of ventilated cavity closures can be used to design a control strategy to form a stable cavity of a given length with minimum gas injection. Furthermore, it allows the maintenance of the cavities at lower ventilation demands, making partial cavity drag reduction and aeration more efficient and sustainable. The rate of increase/decrease of gas ventilation in the ventilation strategy could play a significant role in closure formation. Hence, it is possible that a higher ramp rate can be used to achieve a ventilated cavity more efficiently, i.e. with less volume of gas. However, this is beyond the scope of the current investigation and can be investigated in the future. Furthermore, we recognise that $Re_{H}$ is a more relevant parameter to characterise the ventilated cavities at higher $Fr$ ($\geq$ 5.7). Hence, the effect of $Fr$ and $Re_{H}$ can be segregated with different wedges and experiments across different scales.   

% ============================ movies ============================ %
\section*{Supplementary movies}
Supplementary movies are available \href{https://drive.google.com/drive/folders/16cF_JeeIDr6bzYMTj6aZeJDjnAmB982R?usp=sharing}{here}

\section*{Acknowledgement}
UUG and CP are funded by the ERC Consolidator Grant No. 725183 `OpaqueFlows'. UUG would like to thank TU Delft for supporting his stay at the University of Michigan for the duration of the experiments. NAL, PJ, SLC and HG are supported by the Office of Naval Research, under program managers Dr. Ki-Han Kim and Dr. Julie Young, Grant Number N00014-21-1-2456. 

\section*{Declaration of interest}
The authors report no conflict of interest.

% ============================ Appendices ============================ %
\appendix 

\section{Gas ejection model for transition of REJ to long cavity}  \label{AppC}
\begin{figure}
    \centering
    \includegraphics[width=0.99\textwidth]{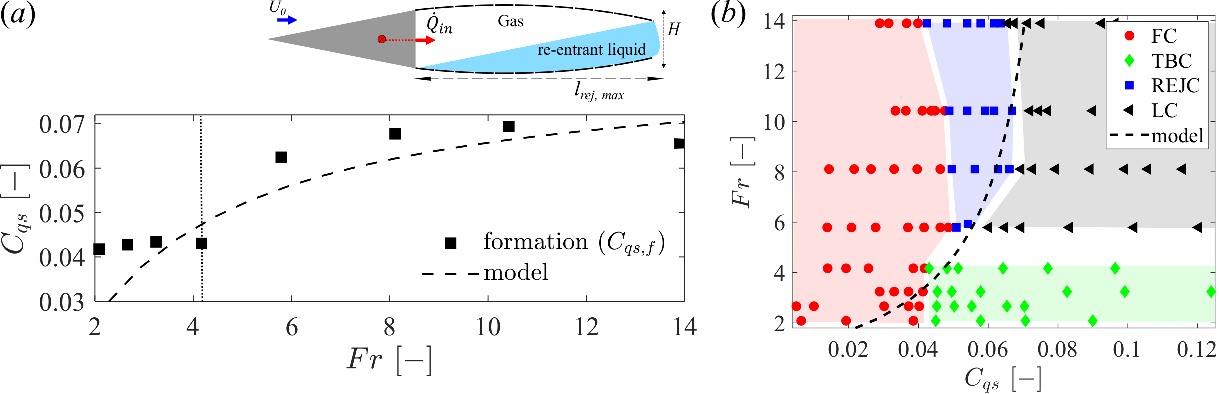}
    \caption{(\textit{a}) $C_{qs,f}$ in comparison to model prediction as a function of $Fr$ (the black vertical dotted line indicates the $Fr$ limit where re-entrant flow dictates transition to a supercavity). The inset shows the schematic of the re-entrant flow. (\textit{b}) Model prediction in the regime map of figure \ref{fig:regime}.}
    \label{form_model}
\end{figure}

The gas ejection rate due to periodically re-entering liquid flow ($\dot Q_{out, rej}$) can be approximated as $V_{rej}f_{rej}$, where $V_{rej}$ and  $f_{rej}$ are the volume and the frequency of the re-entering flow. $V_{rej}$ can then be approximated as $l_{rej}HW/2$ due to the triangular shape of the averaged re-entering flow (see figure \ref{avg_vc}(\textit{c}) and inset in figure \ref{form_model}(\textit{a})). The gas ejection can now be expressed as $C_{qs, out, rej}$ as per equation \ref{cqs} and rearranged to obtain equation \ref{cqsrej1} using $\overline{u_{rej}}/U_0 \approx 0.3$. Finally, we arrive at equation \ref{cqsrej2} using empirical relations obtained experimentally for $l_{rej, max}(Fr)/H$, $St_{H,rej}(C_{qs,in})$ in figure~\ref{fig:rej-spacetime}(\textit{d}) and \ref{unstable_cav}(\textit{e}), respectively. All the symbols have been explained previously.
\begin{equation} \label{cqs}
    C_{qs, out, rej}  = \frac{\dot Q_{out, rej}}{U_{0}WH}
\end{equation}
\begin{equation} \label{cqsrej1}
    C_{qs, out, rej}  =  l_{rej, max} \frac{ f_{rej}  }{2U_{0}} = 0.15\frac{l_{rej, max}}{H} St_{H,rej}   
\end{equation}
\begin{equation}  \label{cqsrej2}
     C_{qs, out, rej}  = 0.15(0.86 Fr-0.73)(0.2-2.38 C_{qs, in})  
\end{equation}

The locus of $C_{qs, in}$ for which $C_{qs, in} > C_{qs, out, rej}$ is shown by the black dashed line in figure \ref{form_model}. It agrees well with $C_{qs, f}$ at higher $Fr$ ($>$ 4.17), suggesting that gas ejection, i.e. displaced due to re-entrant flow accounts for the majority of the gas ejection, preventing the ventilated cavity from growing longer to an LC. Thus, if $C_{qs, in} > C_{qs, out, rej}$, the cavity will experience net growth to an LC from a REJ cavity as it can overcome the gas ejection due to the re-entering liquid flow. Furthermore, at low $Fr$ ($< 4.17$), the predicted ejection due to the re-entrant flow is less than the measured ejection ($C_{qs,f}$). This suggests that the re-entrant flow in such cavities is mild and does not account for the majority of gas leakage. Instead, wave-type instability and the spanwise vortex ejection cause the majority of the gas leakage, consistent with our experimental observations.\par

\bibliographystyle{jfm}
%\bibliography{references11}

\end{document}